\documentclass[twocolumn,superscriptaddress,showpacs,prd,aps,amsmath,amssymb,nofootinbib]{revtex4-1}
\usepackage{graphicx}
\usepackage{longtable}
\usepackage{orcidlink}
\usepackage{color}
\usepackage{makecell}
\usepackage{appendix}
\usepackage{mathrsfs}

\usepackage[normalem]{ulem}
\usepackage{epstopdf}
\usepackage{color}
\usepackage{soul}
\usepackage{ulem}
\usepackage{bm}
\usepackage{xspace}

\usepackage[]{amsmath}
\usepackage[]{amsfonts}
\usepackage[]{amssymb}
\usepackage[]{mathrsfs}
\usepackage[]{mathtools}
\usepackage{times}
\usepackage{hyperref}
\hypersetup{
  colorlinks=true,        
  linkcolor=blue,         
  citecolor=cyan,         
}
\newcommand{\apjl}{Astrophys. J. Lett.}

\newcommand{\ASAS}{Astron. Astrophys.} 
\newcommand{\ASS}{Astrophys. Space Sci.}

\newcommand{\NPB}{Nucl. Phys. B}

\newcommand{\PLB}{Phys. Lett. B}
\newcommand{\PR}{Phys. Rept.} 

\newcommand{\PREV}{Phys. Rev.}

\newcommand{\PTP}{Prog. Theor. Phys.}

\newcommand{\dirac}{\partial\llap{$\diagup$\kern-2pt}}

\newcommand{\fettu}[1]{\mathbf{#1}}

\newcommand{\diag}{\mathrm{diag}}

\newcommand{\Tr}{\mathrm{Tr}}

\newcommand{\e}{\mathrm{e}}

\def\QEQ{{%
			\setbox0\hbox{$I$}%
			\rlap{\hbox to \wd0{\hss--\hss}}\box0
		}}

\begin{document}

\title{Compact Stars as Hideouts For Color-spin-locked Quark Matter: Implications for Powering High-energy Electromagnetic Emissions}

\author{Xin-Ying Song\,\orcidlink{0000-0002-2176-8778}}
\email{songxy@ihep.ac.cn}
\affiliation{University of Chinese Academy of Sciences, Chinese Academy of Sciences, Beijing 100049, China}
\affiliation{Key Laboratory of Particle Astrophysics, Institute of high-energy Physics, Chinese Academy of Sciences, Beijing 100049, China}
\date{\today}

\begin{abstract}
The possibility of compact stars as hideouts for color-spin-locked (CSL) quark matter (QM) is investigated in both MIT bag model and Nambu-Jona-Lasinio (NJL) model. Within the framework of NJL model, the idea of absolutely stable quark matter and the existence of conventional pure quark star (QS) are not supported; in addition, there appears to be no stable hybrid configuration above $2M_\odot$ as the hideout for CSL QM. The stable configurations of massive strange quark stars could be reproduced in the MIT bag model with QCD corrections being taken into account; moreover, they could act as the hiding place for the CSL QM. An interesting scenario is proposed that the phase transition to the CSL phase could occur in the cooling process. The CSL quark matter is an electromagnetic (EM) superconductor of Type-I, and a complete Meissner effect is expected. However, the analysis for this sizable superconductor indicates that most of the magnetic field is frozen inside the quark core with a critical strength, while in some special cases a small fraction could be expelled from a thin layer near the surface in a short time. The analysis on energetics and time scale suggests that this process could act as an inducement mechanism to power typical fast radio bursts, but as a single source of energy, it is unlikely to generate other EM emissions such as gamma-ray bursts and giant flares.

\end{abstract}

\maketitle

\section{Introduction}
It is most possible that the quark matter (QM) exists in the core of compact stars, as quarks are deconfined in extremely dense baryon matter, and the strange quark matter is predicted to be the ground state of QCD at finite baryon number~\cite[][]{1971PhRvD...4.1601B,1975PhRvL..34.1353C,1976PhLB...62..241B,1984PhRvD..30..272W, 1984PhRvD..30.2379F,1986ApJ...310..261A}.
In sufficiently cold and dense quark matter, an attractive interaction provided by the gluon exchange, no matter how weak, results in the formation of a quark Cooper pair condensate~\cite[][]{1977NuPhB.129..390B,1999NuPhB.537..443A}. This phenomenon is color superconductivity, which is similar to the electronic superconductivity caused by the attractive interaction between electrons mediated by the exchange of virtual phonons in certain metals and alloys (i.e. BCS theory~\cite[][]{1957PhRv..108.1175B}).

The quark pairing in terms of representations of the color gauge group is formulated as the color gauge group SU(3)$_c$, i.e. $[3]_{c} \otimes[3]_{c}=[\overline{3}] ^A_{c}\oplus[6]^S_{c}$, where the number of colors is three, $N_{c} = 3$. The quarks interact in an antisymmetric ($A$) anti-triplet channel ($[\overline{3}]^A_{c}$) and a symmetric ($S$) sextet channel($[6]^S_{c}$) which are attractive and repulsive, respectively. The spin-zero condensate requires an antisymmetric anti-triplet channel in the flavor ($f$) structure ($[\overline{3}]^A_{f}$),  since the overall wave function of the Cooper pair has to be antisymmetric, which leads to even-parity, spin-singlet pairing. In cold and very dense three-flavor quark matter, the color-flavor locked (CFL) phase is the ground state~\cite{1999PhLB..470..189S,1999NuPhB.537..443A,2000NuPhB.575..269S,2001PhRvL..86.3492R}. As the density drops below some critical value of the strange quark mass, it is replaced by the gapless CFL
(gCFL) phase~\cite[][]{2004PhRvL..92v2001A,2005PhRvD..71e4009A}. 

However, as the density continues to decrease, the quark chemical potential ($\mu$)\footnote{$\mu$ is related to the baryon chemical potential by $\mu_{\rm B}=3\mu$.} is not sufficient to populate any strange quarks, the quark matter should be two-flavor. Thus, some two-flavor condensates are proposed, such the standard two-flavor color-superconducting (2SC) phase, the gapless 2SC (g2SC) phase, as well as unpaired normal quark (NQ) phase. Some theoretical investigations~\citep[][]{2005PhR...407..205B,  2005PhRvD..72f5020B, 2005PhRvD..72c4004R} have been performed with self-consistent treatment of quark masses to gain the information at moderate chemical potentials. One of the conclusions is that the pure 2SC condensate does not occur for weak and intermediate diquark coupling at low temperatures (less than $10$ MeV) while the NQ phase (or the mixed phase of 2SC and NQ) is most favored~\citep{2005PhRvD..72c4004R}. 

In both CFL and 2SC phases the Fermi momenta of the quark flavors participating in pairing are assumed to be equal. This is a necessary condition for the conventional BCS pairing mechanism. However, for moderate densities ($\mu\sim400$ MeV), this assumption could be not valid; therefore, the ground state is neither the (pure) CFL nor the (pure) 2SC state~\cite[]{2005PhRvD..71e4016S}. 
Some other possibilities have been proposed, such as Larkin-Ovchinnikov-Fulde-Ferrell (LOFF) phase~\cite[][]{1964PhRv..135..550F,1969JETP...28.1200L} and single-flavor pairing. The spin-one state corresponds to the spin-triplet ($[3]^S_J$), thus, the flavor structure ($[6]^S_{f}$) of          color-spin-locked (CSL) phase allows pairing in single-flavor\footnote{In this paper, other spin-one phases (e.g., the polar, A and planar phases) are not considered.}. The gap parameters ($\Delta$) of CSL phase have an order of 10 keV to 1 MeV, which are small compared with those from 2SC and CFL~\cite[]{2005PhRvD..71e4016S}.

The electromagnetic (EM) Meissner effect is a matter of debate for the color superconductor (CSC)~\cite[][]{2000NuPhB.571..269A,2003PhRvL..91x2301S}. 
There is almost not EM Meissner effect in spin-zero CSC as concluded in some works (e.g., see~\cite[][]{2000NuPhB.571..269A}). In contrast, the CSL condensate is a Type-I EM superconductor and the magnetic field would be expelled from macroscopic regions if it
does not exceed the critical magnetic field ($B_{\rm c}$)~\citep{2003PhRvL..91x2301S}. If a quark star (QS) or a quark core of a hybrid star could be a hideout for QM in CSL phase, once the temperature falls below the critical value ($T_{\rm c}$), the phase transition to the CSL state would affect the properties of the compact star and may lead to observable consequences. It is interesting to investigate if it could power high-energy EM emissions, such as fast-radio bursts (FRBs), a magnetar giant flares (MGF) of soft gamma-ray repeaters (SGR), and gamma-ray bursts (GRBs).

The paper is organized as follows: in Section~\ref{sec:modelandpara}, the Nambu-Jona-Lasinio (NJL) model and MIT bag model are used to describe properties of dense quark matter and color superconducting phases; the possibilities of compact stars as hideouts for CSL QM within the frameworks of two models are discussed respectively; in Section~\ref{sec:QS},  the expulsion of magnetic flux induced by the forming of CSL condensate is described; the possible observable consequences are discussed and summarized in Section~\ref{sec:discussion}.

\section{The model for quark matter }\label{sec:modelandpara}

\subsection{NJL model}
The locally neutral three-flavor quark matter is studied within the framework of Nambu–Jona-Lasinio (NJL) model~\protect\cite{PhysRev.122.345,PhysRev.124.246, 2005PhR...407..205B}. The constituent quark masses are dynamically generated and treated self-consistently. The mean-field Lagrangian density for quarks is given by
\begin{equation}
    \mathscr L^{\rm MF}=  \mathscr L_0 +\mathscr L_{\rm int} ,
\end{equation}
where the $\mathscr L_0 = \bar \psi ( i \dirac -\hat{m}) \psi $ is the free part and quark spinor field $\psi_{f}^c$ carries color ($c=r$, $g$, $b$) and flavor ($f=u$, $d$, $s$) indices; the matrix of quark 
current masses is given by $\hat{m} = \diag_{f}(m_u$, $m_d$, $m_s)$.
$\mathscr L_{\rm int}= \mathscr L_{q\overline{q}} + \mathscr L_{qq} + \mathscr{L}_{\rm 't~hooft}$, is the interaction part that includes the quark-antiquark ($q\overline{q}$), diquark ($qq$) and 't Hooft interaction terms. The $q\overline{q}$ term is given by,
\begin{equation}
    \mathscr L_{q\overline{q}}= G_S\sum\limits_{a=0}\limits^{8} {(\bar \psi \lambda_a \psi)^2+(\bar \psi i \gamma_5 \lambda_a \psi)^2 }  ,\\
\end{equation}  
while the 't Hooft interaction term is given by,
\begin{equation} 
    \mathscr L_{\rm 't~hooft}=-K \{ \det_{f}\left[ \bar \psi \left( 1 + \gamma_5 \right) \psi
\right] + \det_{f}\left[ \bar \psi \left( 1 - \gamma_5 \right) \psi
\right]\},
\label{eq:Lqq_thooft}
\end{equation}
where $G_S$ and $K$ are constants for $q\overline{q}$ coupling and 't Hooft interaction respectively. $\lambda_a$ with $a=1,\ldots,8$ are
the Gell-Mann matrices in flavor space, and $\lambda_0\equiv
\sqrt{2/3} \,\openone_{f}$.
For $\mathscr L_{qq}$, the spin-0 condensates 
have different diquark pairing channels from those of spin-1 condensates. For spin-0 cases (the Greek indices signify flavors and the Latin indices signify colors, the same below):

\begin{small}
\begin{equation}
    \mathscr L_{qq, \rm{spin=0}}=G_D \sum_{\gamma,c} \left[\bar{\psi}_{\alpha}^{a} i \gamma_5
\epsilon^{\alpha \beta \gamma}
\epsilon_{abc} (\psi_C)_{\beta}^{b} \right] \left[ 
(\bar{\psi}_C)_{\rho}^{r} i \gamma_5
\epsilon^{\rho \sigma \gamma} \epsilon_{rsc} \psi_{\sigma}^{s} 
\right],\\
\label{eq:Lqq_s0}
\end{equation}
\end{small}
where the $G_D$ is diquark coupling ($G_D= \frac{3}{4} G_S$ denotes an intermediate diquark coupling strength, while $G_D= G_S$ denotes a strong one); the charge conjugate spinors are defined as follows: $\psi_C = C \bar \psi^T$ and $\bar \psi_C = \psi^T C$, where $\bar\psi=\psi^\dagger \gamma^0$ is the Dirac conjugate spinor and $C=i\gamma^2 \gamma^0$ is the charge conjugation matrix. For the CSL case:
\begin{equation}    
    \mathscr L_{qq, \rm{CSL}}=H \sum_{f=u,d,s}
    \left[ 
\bar{\psi}_f i \gamma_\nu
\lambda_A \psi_{C} \right]\left[\bar{\psi}_{C, f} i \gamma_\nu
\lambda_{A}\psi_{f}\right]
 ,\\
\label{eq:CSL}
\end{equation}
where $H=\frac{3}{8} G_S$; $(\nu, A)=(3,2)$, $(1,7)$ and $(2,5)$.

The grand partition function, up to an irrelevant normalization
constant, is given by 
\begin{equation}
\label{eq:Z}
\mathcal{Z} \equiv \e^{-\Omega V/T}
= \int \mathcal{D} \bar\psi \mathcal{D} \psi \, \e^{i
\int_X \left( \mathcal{L} + \bar\psi \hat{\mu} \gamma^0 \psi
\right) } \; ,
\end{equation}
where $\Omega$ is the thermodynamic potential density, $V$ is the 
volume of the three-space, and $\hat\mu$ is a diagonal matrix of 
quark chemical potentials  which will be specified in different cases, $
\mu_{ab}^{\alpha\beta} = \left(
  \mu \delta^{\alpha\beta}
+ \mu_Q Q_{f}^{\alpha\beta} \right)\delta_{ab}
+ \left[ \mu_3 \left(T_3\right)_{ab}
+ \mu_8 \left(T_8\right)_{ab} \right] \delta^{\alpha\beta} \; $. In chemical equilibrium, the nontrivial components of this matrix are extracted from the following relation:  $\mu_Q$ is the chemical potential of electric charge (in the following, $\mu_e=-\mu_Q$), while $\mu_3$ and $\mu_8$ are color chemical potentials associated  with two mutually commuting color charges of the SU(3)$_c$ gauge group. The explicit form of the electric charge matrix is $Q_{f}=\mbox{diag}_{f}(\frac23,-\frac13,-\frac13)$, and the explicit form of the color charge matrices are  $T_3=\mbox{diag}_c(\frac12,-\frac12,0)$ and  $\sqrt{3}T_8=\mbox{diag}_c(\frac12,\frac12,-1)$.

\begin{figure*}
\begin{center}
 \centering
 \includegraphics[width=0.5\textwidth]{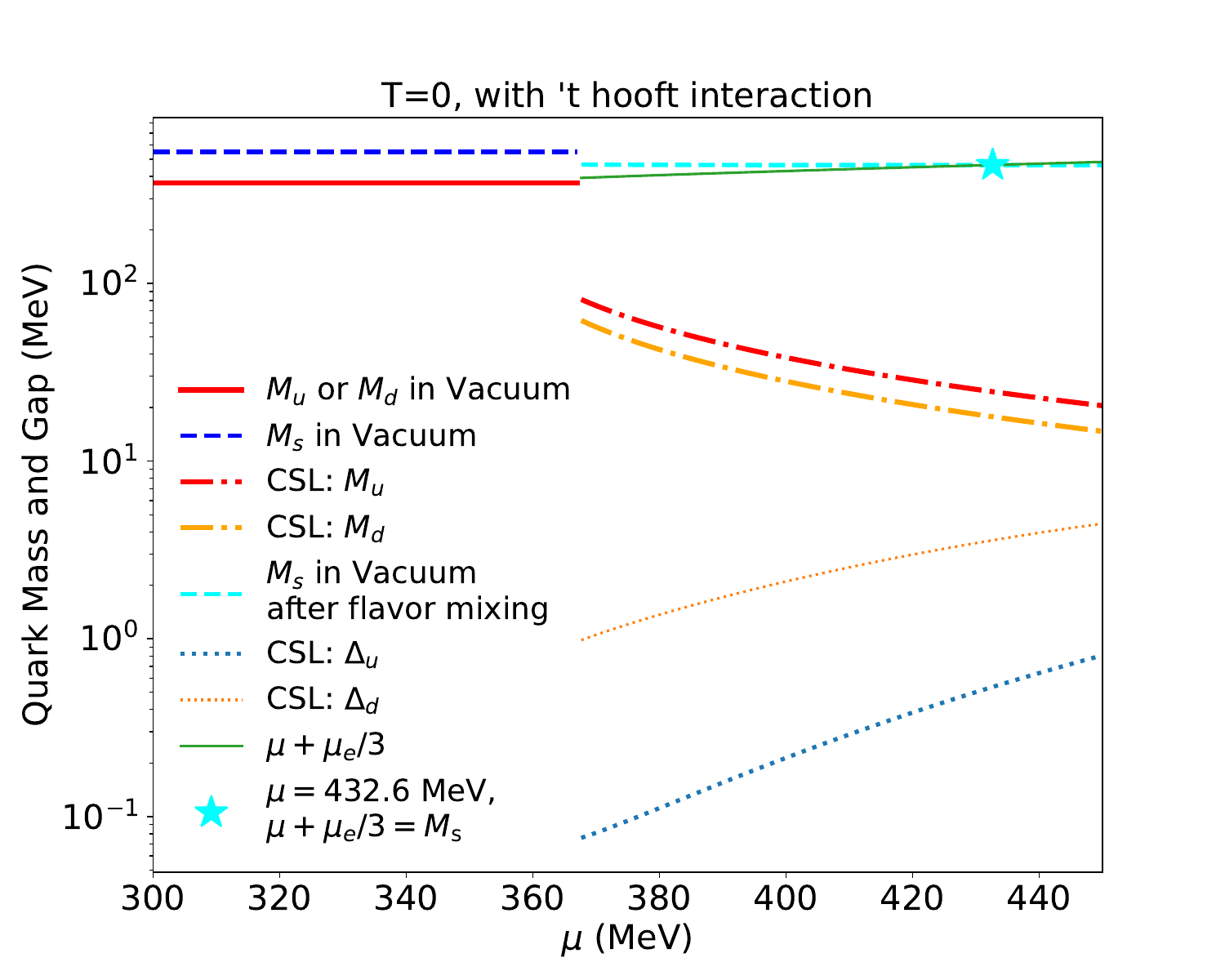}\put(-80,140){(a)}
 \includegraphics[width=0.5\textwidth]{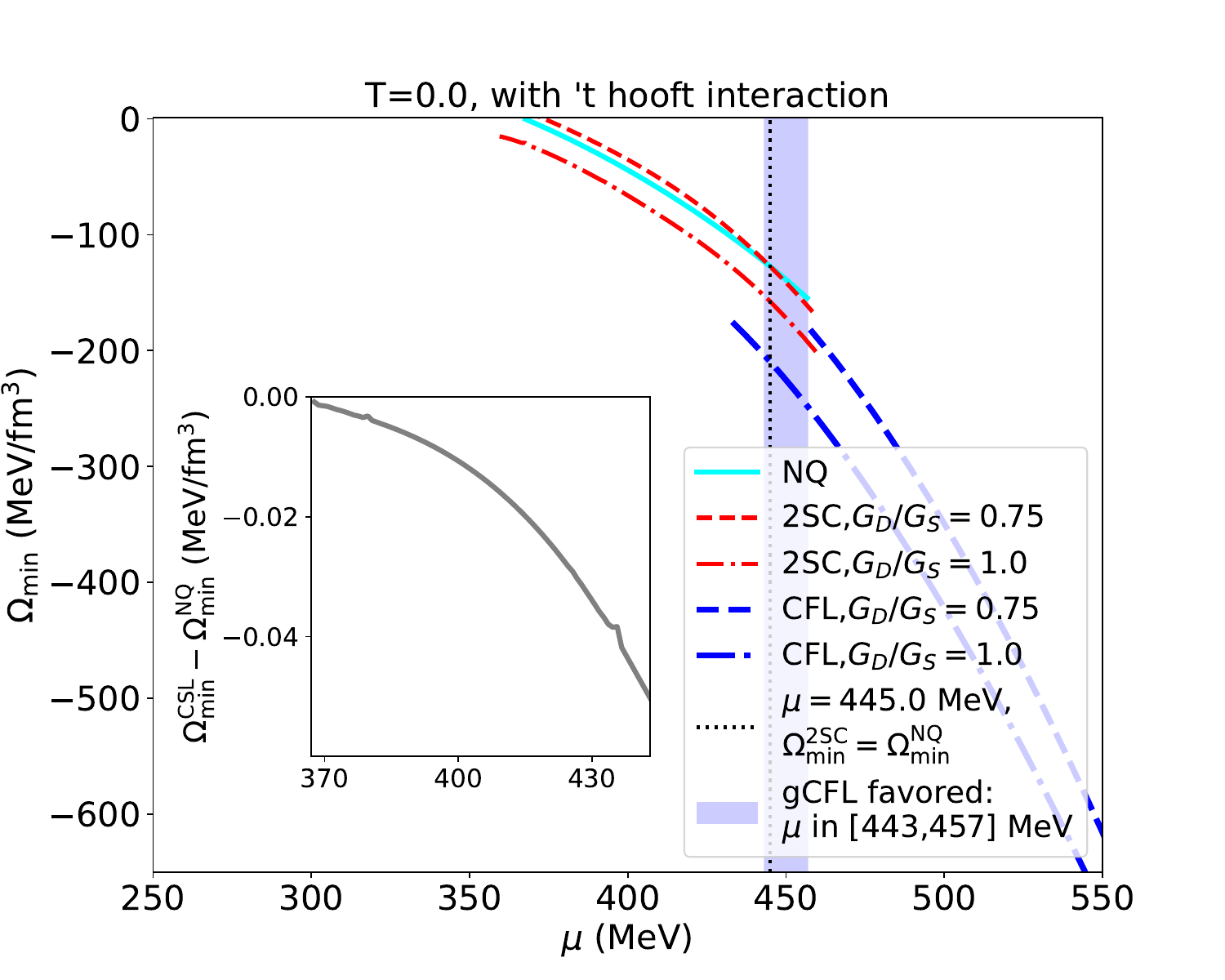}\put(-80,140){(b)}\\
 \includegraphics[width=0.5\textwidth]{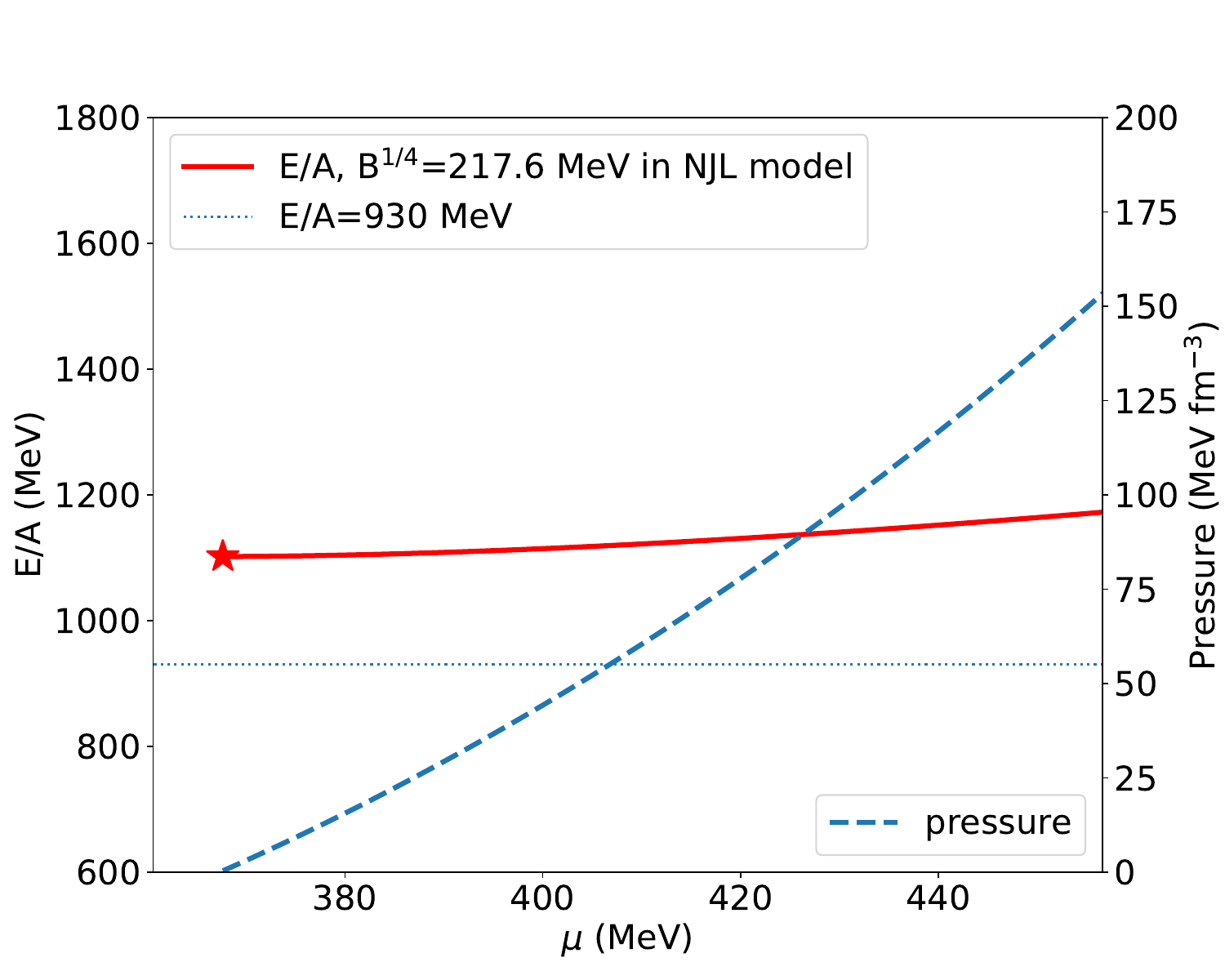}\put(-70,165){(c)}
 \includegraphics[width=0.5\textwidth]{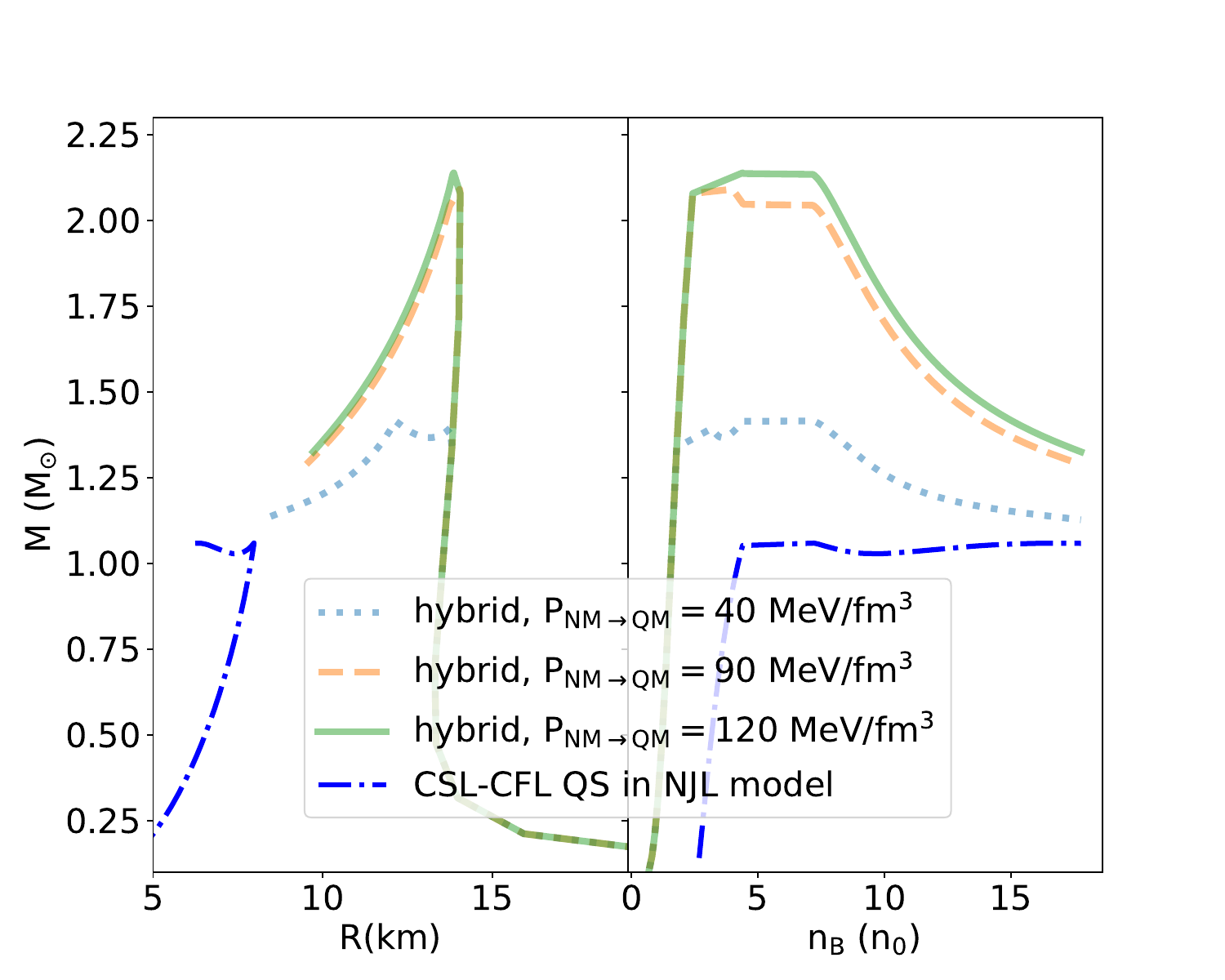}\put(-210,165){(d)}\\
\caption{ (a) The dynamic quark masses and gaps in CSL phase. The blue dashed line denotes $M_{s, \rm Vac}=549$ MeV in vacuum, while the cyan dashed line denotes $M_{s, \rm Vac}$ around 464 MeV in vacuum due to the flavor mixing with considering 't hooft interaction. The dot-dashed lines denote the dynamic masses of light quarks, while the dotted lines denote the gaps. The green thin solid line denotes the values of $\mu_s=\mu+\mu_e/3$ varying with $\mu$, with a cyan star denoting the point of $\mu_s=M_s$. (b) Minima of the thermodynamical potential
for neutral three-flavor quark matter at $T=0$ as a function of the quark chemical potential. The red dashed and dot-dashed lines denote $\Omega_{\rm min}$ in 2SC phase with $G_D/G_S=0.75$ and $G_D/G_S=1.0$, while the blue ones denote those of CFL phases. The solid cyan line denotes $\Omega_{\rm min,NQ}$. The blue shadow denotes the region that the gCFL phase is most favored. The dotted vertical line ($\mu=445$ MeV) denotes the beginning that the 2SC phase becomes more stable than NQ and CSL phases. 
(c) $E/A$ of the CSL QM in the NJL model. (d) Sequences of pure QSs and hybrid stars (neutron star matter is describe by the Walecka model~\citep[][]{1997csnp.book.....G}). $n_{\rm B}$ is the central density with the unit of $n_0$. Note that the `plateau' from $4.4n_0$ to $7.2n_0$  in the right panel is not real because this is discontinuous region in density from CSL to CFL phase.}\label{fig:MD}
\end{center}
\end{figure*}

The interaction terms are linearized
in the presence of the diquark condensates ($\Delta_{c, \rm{spin=0}}\propto
(\bar{\psi}_C)_{\alpha}^{a} i \gamma_5 \epsilon^{\alpha \beta c} 
\epsilon_{a b c} \psi_{\beta}^{b}$ for spin-0 cases and $\Delta_{f, \rm{CSL}}\propto\bar{\psi}_f i \gamma_\nu
\lambda_A \psi_{C}$ for the CSL state) and 
quark-antiquark condensates ($\sigma_\alpha \sim \bar \psi_\alpha^a 
\psi_\alpha^a$). The thermal potential for the case of spin-0 is given by\footnote{Two general formulae are used in the derivation: $\int d^Dx~ e^{-\frac{1}{2} x \hat{A} x}= (2\pi)^{\frac{D}{2}} (\det \hat{A})^{\frac{1}{2}}$; $\ln\det|\hat{A}|= \Tr (\ln \hat {A})$ (see \cite{1976PThPh..56..947K}).}, 
\begin{eqnarray}
\Omega_{\rm{spin=0}} &=& \Omega_{e}-\Omega_{\rm Vac} +
\frac{1}{4 G_D} \sum_{i=1}^{3} \left| \Delta_i \right|^2
+2 G_S \sum_{\alpha=1}^{3} \sigma_\alpha^2 \nonumber\\
&-& 4 K \sigma_u \sigma_d \sigma_s
-T \sum_{n} \int_{|\fettu k|<\Lambda} \frac{d^3 k}{(2\pi)^3} \frac{1}{2} \Tr \ln  \frac{S^{-1}(i\omega_n, \fettu{k})}{T}, \nonumber \\
\label{eq:Omega_s0}
\end{eqnarray}
while that for CSL phase is,
\begin{eqnarray}
\label{eq:omega_s1}
\Omega_{\rm{CSL}} &=& \Omega_{e} -\Omega_{\rm Vac} +
\frac{1}{H} \sum_{f=u,d,s} \left| \Delta_f \right|^2
+2 G_S \sum_{\alpha=1}^{3} \sigma_\alpha^2 \nonumber\\
&-& 4 K \sigma_u \sigma_d \sigma_s
- T \sum_{n} \int_{|\fettu k|<\Lambda} \frac{d^3 k}{(2\pi)^3} \frac{1}{2} \Tr \ln  \frac{S_f^{-1}(i\omega_n, \fettu{k})}{T} , \nonumber \\
\end{eqnarray}
where $\Omega_e$ is the thermodynamic potential of ultrarelativistic
electrons. $\Omega_{\rm Vac}$ is contribution from the vacuum at $T=\mu=0$, i.e. the bag constant; it is the thermodynamic potential of nontrivial vacuum which is dynamically generated in NJL model, and has been subtracted in order to get zero pressure in vacuum. In real QCD the ultraviolet modes decouple because of asymptotic freedom, but in the NJL model this feature is added by hand, through a UV momentum cutoff $\Lambda$ in the momentum integrals.
In Equation (\ref{eq:Omega_s0}), $S^{-1}$ is the inverse full quark propagator 
in the Nambu-Gorkov representation, 
\begin{equation}
S^{-1} = 
\left(
\begin{array}{cc}
[ G_0^+ ]^{-1} & \Phi^- \\
\Phi^+ & [ G_0^- ]^{-1}
\end{array}
\right) \; ,
\label{off-d}
\end{equation}
with the diagonal elements being the inverse Dirac propagators of quarks and of charge-conjugate quarks, 
\begin{equation}
[ G_0^\pm ]^{-1} = \gamma^\nu k_\nu 
\pm \hat \mu \gamma_0 - \hat{M} \; ,
\end{equation}
where $k^\mu = (k_0, \fettu{k})$ denotes the four-momentum of the quark. $\hat{M}=\mbox{diag}_f(M_u, M_d, M_s)$ is the constituent mass matrix, and defined as,
\begin{equation}
M_\alpha= m_\alpha -4G_S\sigma_\alpha+2K\sigma_\beta \sigma_\gamma,
\end{equation}
which is treated self-consistently as dynamically generated quantity. The definition of $S^{-1}_f$ in Equation~(\ref{eq:omega_s1}) and corresponding derivation are similar. The details of evaluations for $\Omega_{\rm spin=0}$ and $\Omega_{\rm CSL}$ are presented in Appendix~\ref{sec:dispersion}.
The set of parameters are listed as below~\citep{1996PhRvC..53..410R},
\begin{subequations}
\label{model-parameters}
\begin{eqnarray}
m_{u,d} &=& 5.5 \; \mathrm{MeV} \; , \\
m_s &=& 140.7 \; \mathrm{MeV} \; , \\
G_S \Lambda^2 &=& 1.835 \; , \\
K \Lambda^5 &=& 12.36 \; , \\
\Lambda &=& 602.3 \; \mathrm{MeV} \; ,
\label{Lambda}
\end{eqnarray}
\end{subequations}
which could reproduce the masses of mesons of vacuum QCD. The constituent masses in vacuum for light and strange quarks are $368$ MeV and $549$ MeV,  which are the stable solution for $\Omega_{\rm Vac}$, as shown in FIG.~\ref{fig:MD} (a).

The gaps and dynamic quark masses are determined by minimization of the grand canonical thermodynamical
potential as 
\begin{subequations}
\label{eq:gapeqns}
\begin{eqnarray}
\frac{\partial \Omega}{\partial \sigma_\alpha} &=& 0 \;  , \\
\frac{\partial \Omega}{\partial \Delta_{i(f)}} &=& 0 \; .
\end{eqnarray}
\end{subequations}
Since the quark matter should be locally color and electric charge neutral, the color and electric chemical potentials are evaluated by enforcing zero color and electric charge as, 
\begin{subequations}
\label{eq:neutrality}
\begin{eqnarray}
n_e &\equiv& -\frac{ \partial \Omega }{\partial \mu_e} = 0 \; , \\
n_3 &\equiv& -\frac{ \partial \Omega }{\partial \mu_3} = 0 \; , \\
n_8 &\equiv& -\frac{ \partial \Omega }{\partial \mu_8} = 0 \; .
\end{eqnarray}
\end{subequations}
Note that the CSL phase is naturally color neutral, thus it is not necessary to impose color neutrality and one has $\mu_3=\mu_8=0$.
The calculation is performed at $T=0$ to approximately obtain the properties of QM in a cold QS. According to the conclusion of previous work~\citep{2005PhRvD..72c4004R}, in the regime of weak and intermediate diquark coupling strength, the unpaired normal quark (NQ) phase is most favored in the moderately dense matter at $T\lesssim10$ MeV, however, color-superconducting condensates considered in that work are only those of spin-0. The CSL phase could be more favored than the NQ phase at some densities~\citep{2007Ap&SS.308..443A}. Therefore, we should compare these minima of thermodynamical potential to find the most stable phase at different densities. The types and gaps of phases are listed as below,
\begin{itemize} 
    \item 2SC: $\Delta_3>0$, $\Delta_2=\Delta_1=0$; the gaps $\Delta_1$, $\Delta_2$, and $\Delta_3$ correspond to the down-strange, the up-strange and the up-down diquark condensates, respectively;
    \item CSL: each flavor is paired, and the gap is $\Delta_f$; 
    \item NQ: $\Delta_3=\Delta_2=\Delta_1=0$;
     \item (g)CFL: $\Delta_3\gtrsim\Delta_2=\Delta_1$; as $\mu$ decreases,
     one has CFL$\rightarrow$ gCFL, and $\Delta_3>\Delta_2>\Delta_1>0$. 
\end{itemize}

The dynamical quark masses and gaps of CSL condensate in the moderate chemical potentials are shown in FIG.~\ref{fig:MD} (a). When $\mu\geq368$ MeV, the light quarks are populated, and strange quark mass in vacuum falls from 549 MeV to about 464 MeV due to the flavor mixing~\citep{2005PhR...407..205B}. The gaps $\Delta_{u,d}$ range from hundreds of keV to a few MeV. Above $\mu=433$ MeV (denoted by a cyan star as shown in FIG.~\ref{fig:MD} (a)), the value of $\mu+\mu_e/3$ begins to be larger than $M_s$ in vacuum, and the density of strange quarks becomes nonzero. 

Minima of the thermodynamical potential for QM at $T=0$ as a function of $\mu$ are shown in FIG.~\ref{fig:MD} (b). For $G_D/G_S=3/4$, gCFL is most favored at $443$ MeV $\lesssim\mu\lesssim457$ MeV, which is denoted by a narrow blue shadow\footnote{The calculation for gCFL is not performed in this paper; numerical results of gCFL from \cite{2005PhRvD..72c4004R} are used in the FIG.~\ref{fig:MD} (b).}. 
Although the difference between the $\Omega^{\rm CSL}_{\rm min}$ and $\Omega^{\rm NQ}_{\rm min}$ is small as shown in the sub-plot in FIG.~\ref{fig:MD} (b), the CSL phase is more stable than that of NQ at moderate densities. Therefore, the most favored phase for $368$~MeV $<\mu<443$~MeV is CSL.
Note that for $\mu\gtrsim445$ MeV ($\mu=445$ MeV at which $\Omega^{\rm 2SC}_{\rm min}=\Omega^{\rm NQ}_{\rm min}$ is denoted by a dotted vertical line), the 2SC phase begins to be more stable than the NQ phase, which is consistent with the result of \cite{2005PhRvD..72c4004R}.
For $G_D/G_S=1$, 2SC is the most favored phase at the moderate chemical potentials, however, this is not the case that interests us.


The quark matter is absolutely stable if the energy per baryon ($E/A$) is less than that of $^{56}$Fe, i.e. $E/A<930$ MeV. In fact, the NJL model does not support the idea of absolutely stable quark matter if one must keep the vacuum properties of the model at least qualitatively unchanged~\citep[][]{2005PhR...407..205B}. The bag constant is determined to be $B^{1/4}=218$ MeV ($B=292$ MeV fm$^{-3}$) with parameters in ~\citep{1996PhRvC..53..410R} in NJL model, and $E/A$ at zero-pressure for CSL phase is larger than 930 MeV as shown in Figure~\ref{fig:MD} (c). Thus, the possibility of hybrid configurations should be considered. 

 The critical pressure from nuclear matter to quark matter is unknown, and three values ($P_{\rm NM\rightarrow QM}= 40$, 90, 120 MeV fm$^{-3}$) are used in the calculation.
 The corresponding sequences for hybrid configurations are obtained by solving the well-known Tolman-Oppenheimer-Volkoff (TOV) equation for hydrostatic equilibrium of self-gravitating matter~\citep{1939PhRv...55..374O}, as shown in Figure~\ref{fig:MD} (d). The maximum mass of static configurations ($M_{\rm max}^{\rm static}$) for hybrid stars could reach up to above $2M_{\odot}$ if $P_{\rm NM\rightarrow QM}\gtrsim 90$ MeV fm$^{-3}$, which is compatible with the observed astrophysical data~(e.g., see \citep[][]{2012ARNPS..62..485L}). For the conventional QS constructed in NJL model, $M_{\rm max}^{\rm static}\sim1.06 M_{\odot}$, that is much smaller. 
 However, this hybrid configuration become unstable with respect to the gravitational collapse once quark matter is formed above $2M_{\odot}$ ($dM^{\rm static}/d\rho_{\rm c}<0$, where $M^{\rm stat}$ is the gravitational mass of the static compact star and $\rho_{\rm c}$ is the central energy density), thus there are not stable hybrid stars as hideouts for CSL QM with $M^{\rm static}_{\rm max}>2M_{\odot}$.

\subsection{MIT bag model}
The three-flavor strange quark matter is predicted to be the ground state of QCD at finite baryon number within the framework of MIT bag model (e.g. see the references~\cite[][]{1971PhRvD...4.1601B,1976PhLB...62..241B,1984PhRvD..30..272W, 1984PhRvD..30.2379F,1986ApJ...310..261A}.) With QCD corrections being taken into account, the thermodynamic potential density for unpaired QM is (see details in ~\cite{1978PhRvD..17.1109F,1984PhRvD..30.2379F})
\begin{eqnarray}
    \Omega_{\rm NQ} &=& - (1-c)\frac{3 \mu^4}{4\pi^2}
+ \frac{(3-c) M_s^2 \mu^2}{4\pi^2}+ O(\frac{M_s^4}{10^2})\nonumber\\
&+&O(\frac{M_s^6}{10^4\mu^2}) +B.
\label{eq:NQ}
\end{eqnarray}
For free Fermi-gas, $c=0$, while in this analysis the effect of gluon-mediated QCD interactions between the quarks is considered and $c=\frac{2\alpha_{\rm c}}{\pi} >0$ to the order of $O(\alpha_{\rm c})$, where $\alpha_{\rm c}$ is the QCD coupling~\cite{1984PhRvD..30.2379F,1984PhRvD..30.2379F,2001PhRvD..63l1702F}. The masses of light quarks are neglected. The lower limits of $B$ vary with $\alpha_{\rm c}$: $B^{1/4}\geq145$, 137, 128 and 117 MeV for $\alpha_{\rm c}=0$, 0.3, 0.6 and 0.9, which are determined by the lower limit that nuclei with high atomic numbers would be unstable against decay into non-strange two-flavor quark matter~\cite{1984PhRvD..30.2379F}. The thermodynamic potential density of CSL condensate is

\begin{eqnarray}\label{eq:OmegaCSL_MIT}
\Omega_{\rm CSL} &=& \Omega_{\rm NQ}-\frac{\sqrt{2}}{\pi^2}\sum_{i=u, d}\Delta_{i}^2\mu_i^2 \nonumber\\
&-& \frac{\Delta_{s}^2(\mu_s^2-M_s^2)}{\pi^2}\left(\frac{M_s}{\mu_s}+ \sum_{sign=+,-}\sqrt{2sign\sqrt{2}\frac{M_s}{\mu_s}}\right)\nonumber\\
&+&O(\Delta_{i}^4),
\end{eqnarray}

where the second and third terms are the contributions from condensate that are evaluated from the approximation on the Fermi-surface. $\Omega_{\rm CSL}<\Omega_{\rm NQ}$ always works, because the contribution from condensate terms to $\Omega_{\rm CSL}$ is always less than zero. $\Delta_{\rm CSL}$ is taken to be 1 MeV in the numerical calculation.

The thermodynamic potential densities for spin-zero condensates, e.g., 2SC and CFL condensates, are described to be~\cite{2001PhRvD..64g4017A,2002JHEP...06..031A,2005ApJ...629..969A}  
\begin{equation}
    \Omega_{\rm CFL}=\Omega_{\rm NQ}+\frac{3M_s^4-48\Delta_{\rm CFL}^2\mu^2}{16\pi^2}+O(\frac{M_s^6}{\mu^2}),
\end{equation}
 and 
\begin{equation}
    \Omega_{\rm 2SC}=\Omega_{\rm NQ}+\frac{M_s^4-16\Delta_{\rm 2SC}^2\mu^2}{16\pi^2}+O(\frac{M_s^6}{\mu^2}),
\end{equation} 
respectively, where $\Delta_{\rm 2SC}$ denotes the gap of pairs of $u$ and $d$ quarks. Note that the criterion for the stability CFL phase (i.e. $\Omega_{\rm CFL}\leq \Omega_{\rm NQ}$, or the transition point from NQ to CFL phase) is
\begin{equation}\label{eq:CFLcriterion}
  \Delta_{\rm CFL}\geq\frac{M_s^2}{4\mu}  
\end{equation}
 (e.g., see~\cite{2001PhRvD..64g4017A,2002JHEP...06..031A}). Note that the criterion that the 2SC phase is preferred over the NQ phase is the same as in (\ref{eq:CFLcriterion}). Thus, if criterion (19) is valid, the CFL phase is already the most favored since $\Omega_{\rm CFL}<\Omega_{\rm 2SC}$. This is the reason why some works proposed that the 2SC phase may not exist in the three-flavor strange quark matter~\cite{2002JHEP...06..031A}. Since the impact of CSL condensate on the thermodynamic potential density is very small ($\Delta_{\rm CSL}^2/\mu^2\lesssim10^{-5}$) from Eq~(\ref{eq:OmegaCSL_MIT}),  therefore criterion~(\ref{eq:CFLcriterion}) could be taken to be approximately the phase transition point from the CSL phase to CFL phase. The density (or $\mu$) increases from the surface to the core and if with proper values of $M_s$ and $\Delta_{\rm CFL}$, there exists a density (or $\mu$) inside the star above which the criterion of (\ref{eq:CFLcriterion}) begins to be valid; in this case, the QS has a configuration in which a CFL QM core is surrounded by QM in CSL phase; otherwise, if $\Delta_{\rm CFL}$ is large enough, (\ref{eq:CFLcriterion}) is valid for the whole QS, it is a pure CFL QS, which is not of our interest. The values of $\Delta_{\rm CFL}$ range around $10-100$ MeV~\cite{2001PhRvL..86.3492R, 2004PhRvL..92v2001A,2005PhRvD..71e4009A}. Numerical calculations are performed to find if there exists any stable QS with $M_{\rm max}^{\rm stat}>2M_{\odot}$. Furthermore, it should be a hideout for CSL QM; a CFL-QS is not the case of interest to us.
\begin{figure*}
\begin{center}
 \centering
\includegraphics[width=0.5\textwidth]{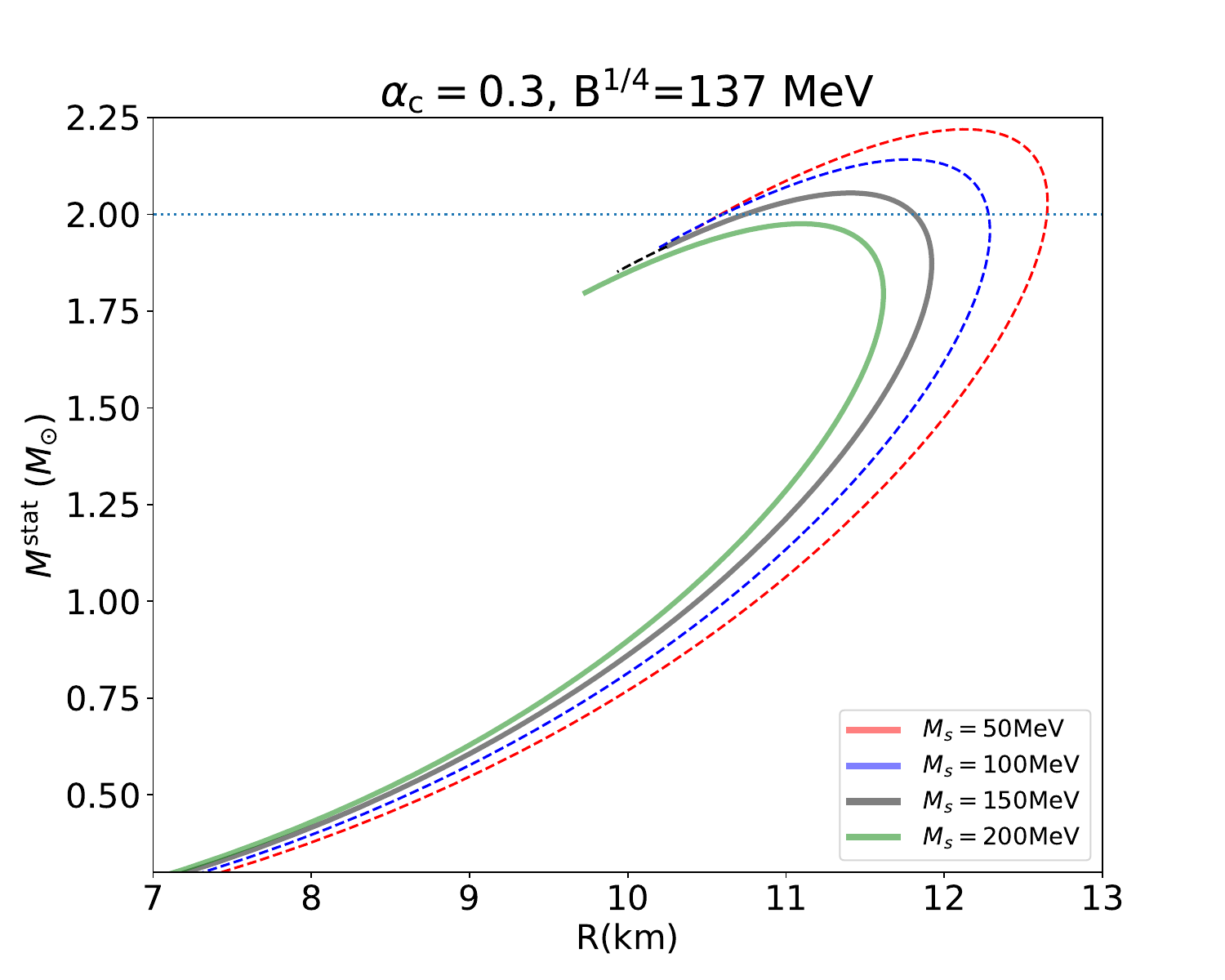}\put(-80,140){(a)}
\includegraphics[width=0.5\textwidth]{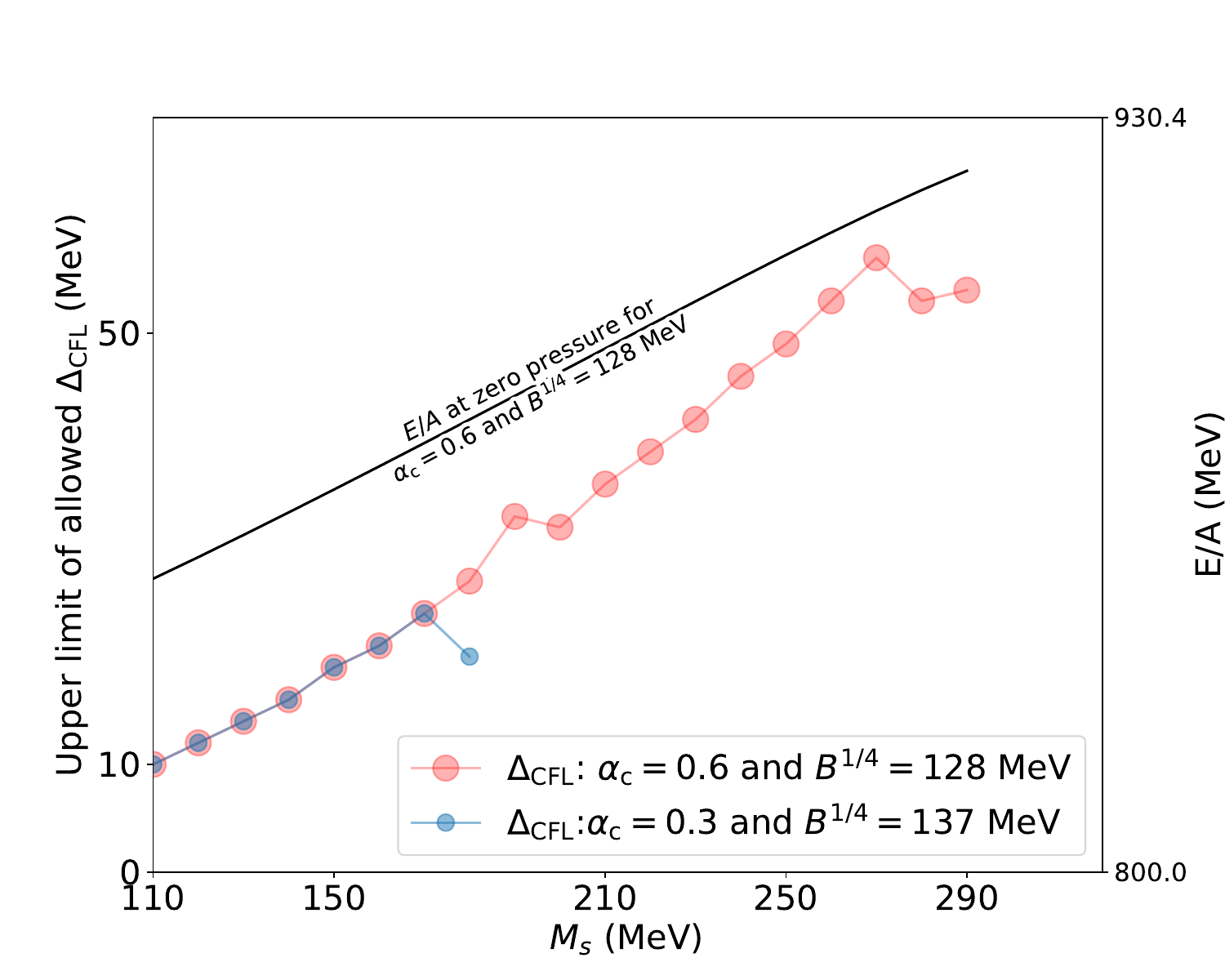}\put(-50,140){(b)}\\
\includegraphics[width=0.5\textwidth]{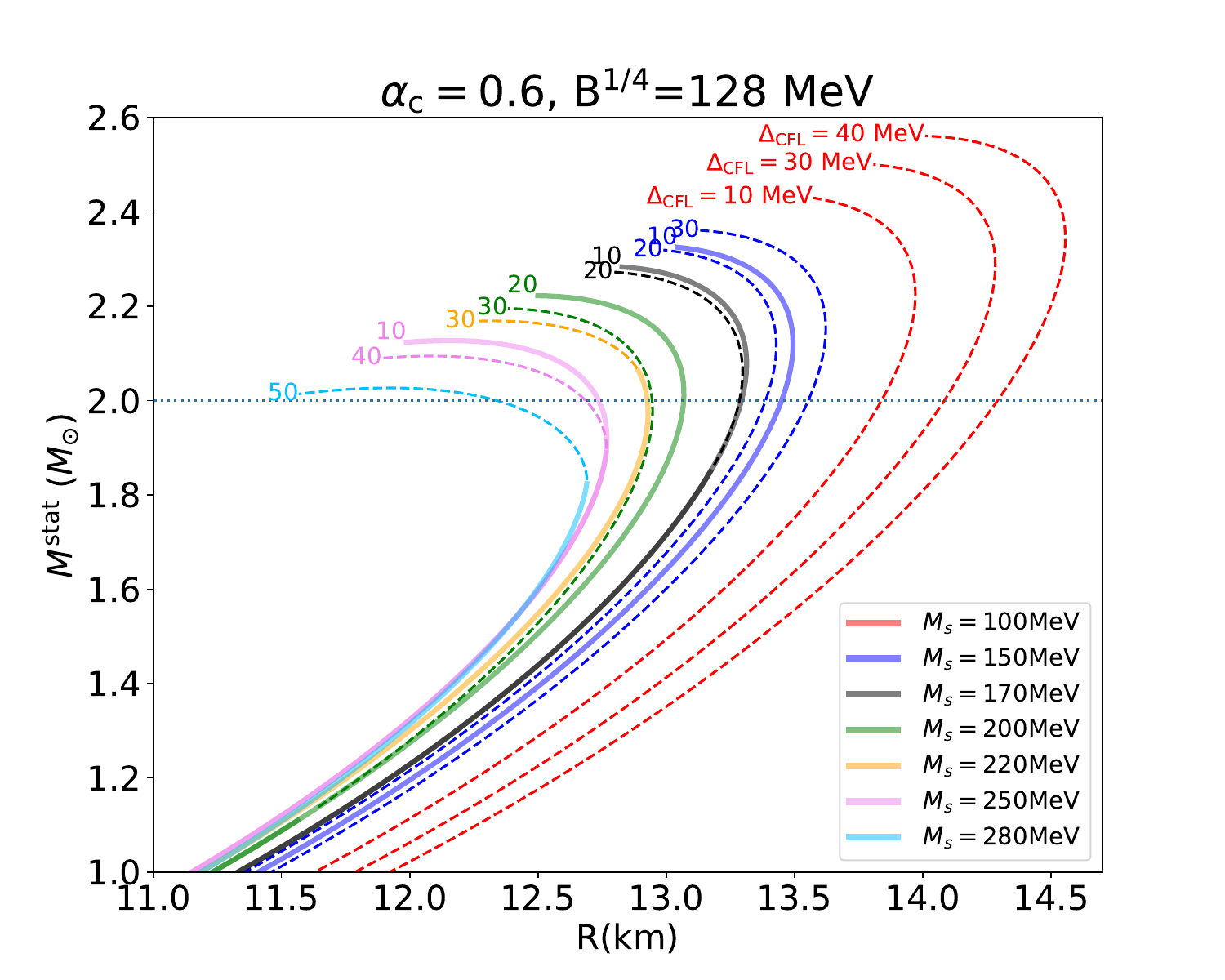}\put(-80,120){(c)}
\includegraphics[width=0.5\textwidth]{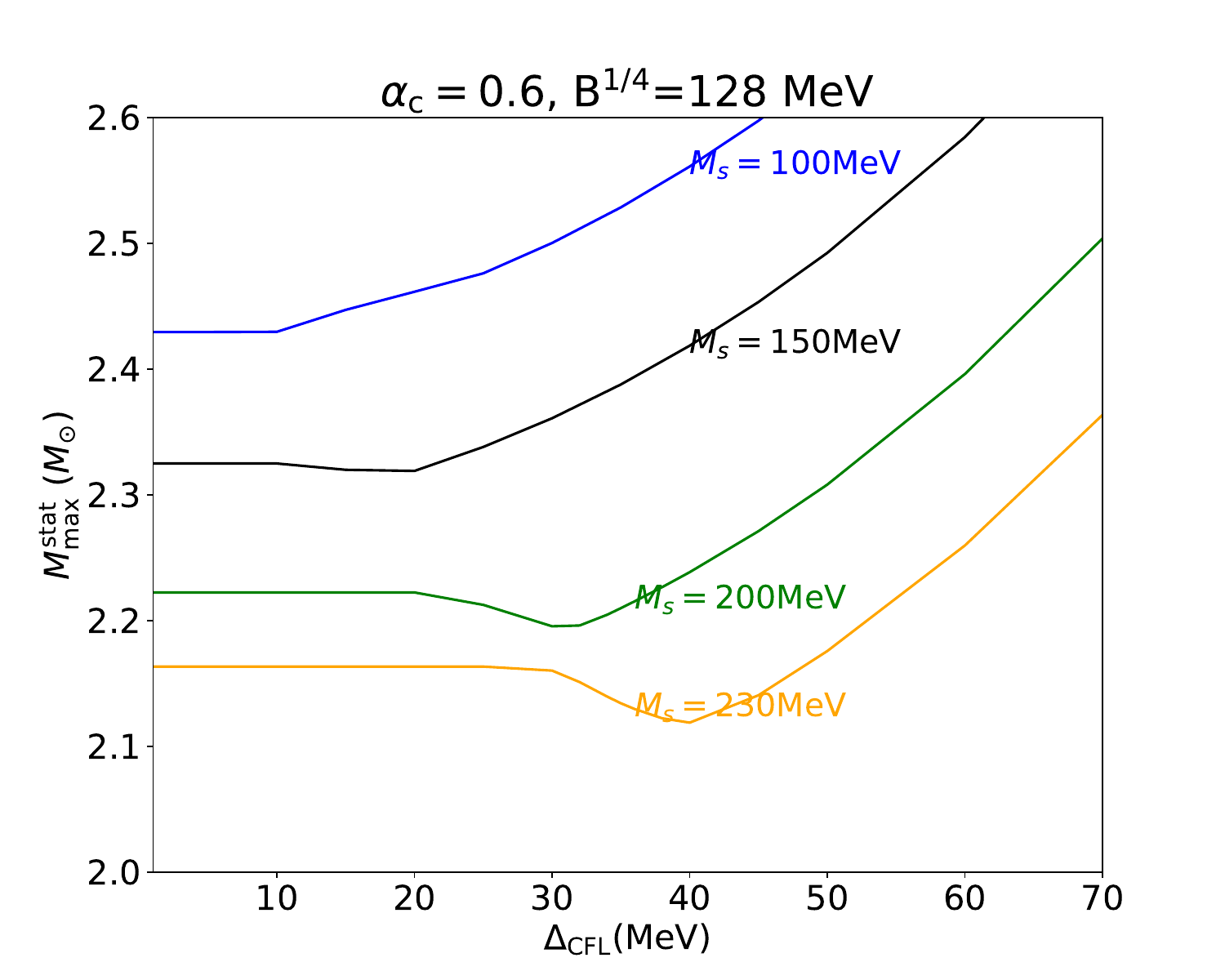}\put(-50,140){(d)}
\caption{ 
(a) and (c) are sequences of strange QS in MIT bag model with $\alpha_{\rm c}=0.3$ and 0.6. Note that different line styles denote different phases in QM. An all-dashed line: QM in CFL phase for the whole star; an all-solid line: QM in CSL phase for the whole star; solid-dashed line denotes a QS with a CFL QM core surrounded by QM in CSL phase in the dashed part ( e.g., $\alpha_{\rm c}=0.6$, $M_s=250$ MeV and $\Delta=40$ MeV). (b) the left Y-axis: the upper limits of allowed $\Delta_{\rm CFL}$ versus $M_s$ for $\alpha_{\rm c}=0.3$ and 0.6. The right Y-axis: $E/A$ at zero pressure as a function of $M_s$ for $\alpha_{\rm c}=0.6$ and $B^{1/4}=128$ MeV. (d) The maximum mass versus $\Delta_{\rm CFL}$ for different $M_s$.
}\label{fig:MITMR}
\end{center}
\end{figure*}

For $\alpha_{\rm c}=0.3$ and $B^{1/4}=137$ MeV, if $M_s\lesssim100$, $\Delta_{\rm CFL}>\frac{M_s^2}{4\mu}$ is satisfied even for a small gap $\Delta_{\rm CFL}=10$ MeV, while for a larger $M_s\gtrsim200$ MeV, $M_{\rm max}^{\rm static}$ can hardly reach above $2M_{\odot}$, as shown in Figure~\ref{fig:MITMR} (a). Therefore, the space of parameters allowed for the existence of CSL matter in QS is narrow, as shown in Figure~\ref{fig:MITMR} (b). For a larger $B^{1/4}=145$ MeV, the case is similar and $M_{\rm max}^{\rm static}$ is smaller. For $\alpha=0.9$, the space of the allowed parameters is even narrower due to the negative baryon electric charge~\cite{1984PhRvD..30.2379F} and will not be calculated or discussed here. 

Some works proposed that $\alpha_{\rm c}\simeq0.6$ could be a reasonable value~\cite{2001PhRvD..63l1702F,2002PhRvD..66j5001A}. For $\alpha_{\rm c}=0.6$ and $B^{1/4}=128$ MeV, $M_s$ could range from 110 to $\sim$300 MeV with $\Delta_{\rm CFL}$ from 10 to more than 50 MeV, as shown in Figures~\ref{fig:MITMR} (b) and (c). $E/A$ at zero pressure increases with $M_s$, but is always below 930.4 MeV. The impact of the formation of CSL condensate on EOS is very small ( $\sim0.1\%$). The effect of CFL condensate formation increases with $\Delta$, which causes a lower $M^{\rm static}_{\rm max}$ (e.g., see the $M-R$ relations of $\Delta_{\rm CFL}=10$ MeV and 40 MeV with $M_s=250$ MeV in Figure~\ref{fig:MITMR} (c)). In fact, the effect of the CFL condensate should be applied more gently, because there might exist a gCFL phase between the CSL and CFL phases; moreover, $\Delta_{\rm CFL}$ is smaller at these moderate densities and increases with density. In summary, the MIT bag model could produce stable and massive SQSs as hiding places for CSL QM. 

Figure~\ref{fig:MITMR} (d) shows how $M_{\rm max}^{\rm stat}$ varies with $\Delta_{\rm CFL}$ for different $M_s$. As shown in Figure~\ref{fig:MITMR} (d), for pure CFL QS, the higher the value of $\Delta_{\rm CFL}$, the larger the maximum mass, which is consistent with the results in \cite{2011ApJ...740L..14W}. For larger $M_s$ ($>100$ MeV), the maximum mass does not increase absolutely monotonically with $\Delta_{\rm CFL}$ at lower $\Delta_{\rm CFL}$ in the configuration in which a CFL QM core is surrounded by QM in the CSL phase. Pure CFL QS is not the topic that this paper focuses on.

\section{The process of expulsion of magnetic flux}\label{sec:QS}
In the hypothesis of a QS as the hiding place for CSL QM, it can be expected that CSL condensate forms during the cooling process. The critical temperature ($T_{\rm c}$) for CSL condensation is of the same order as $\Delta_{\rm CSL}$ ($\simeq1$ MeV). Let us assume a QS with a typical mass of $1.4M_\odot$. If a medium $\Delta_{\rm CFL}\sim50$ MeV is taken, it is required that $M_s\gtrsim230$ MeV with radius of $\sim12$ km from Figure ~\ref{fig:compare}(b) and (c). When the temperature stays above $T_{\rm c}$, the QM (that surrounds the CFL core) remains in the NQ phase. The magnetic field can penetrate the NQ matter, and for simplicity, we assume the magnetic field ($B$) inside the star is uniform.
The magnetic field energy inside star (with the radius $R$) could be estimated as,
\begin{eqnarray}
E_{B} &=&\frac{B^2}{8\pi} \left(\frac{4\pi}{3} R^3\right)\nonumber\\
&=&1.7\times10^{47} {\rm erg} \left(\frac{B}{10^{15}~{\rm G}}\right)^2 \left(\frac{R}{10~ \rm{km}}\right)^3.
\label{eq:EB}
\end{eqnarray}


Once the temperature falls below $T_{\rm c}$, the CSL condensate forms. 
Since the center of quark core has the highest density and $T_{\rm c}$, one may expect that the CSL condensate first forms  in the central of the star during the cooling process\footnote{The cooling of QS is governed by the general relativistic thermal balance, transport properties of the quark matter and the crust properties, which is very complex and not discussed in this work. However, it does not affect our conclusion.}. In a simplified picture, the CSL condensate forms along the edge of NQ matter cooling to $T_{\rm c}$, thus $\tau_{\rm cool}$ could be taken as the time-scale of the CSL front expanding. The consequence of expulsion of magnetic flux during the cooling and phase transition process is affected by the competition between the characteristic time-scale ($\tau_{\rm exp}$) of the expulsion of magnetic flux and $\tau_{\rm cool}$; it is discussed in different cases as follows.

\subsection{\textbf{The 1st Case:}  $\tau_{\rm cool}\ll\tau_{\rm exp}$ }
\begin{figure*}
\begin{center}
 \centering
  \includegraphics[width=.9\textwidth]{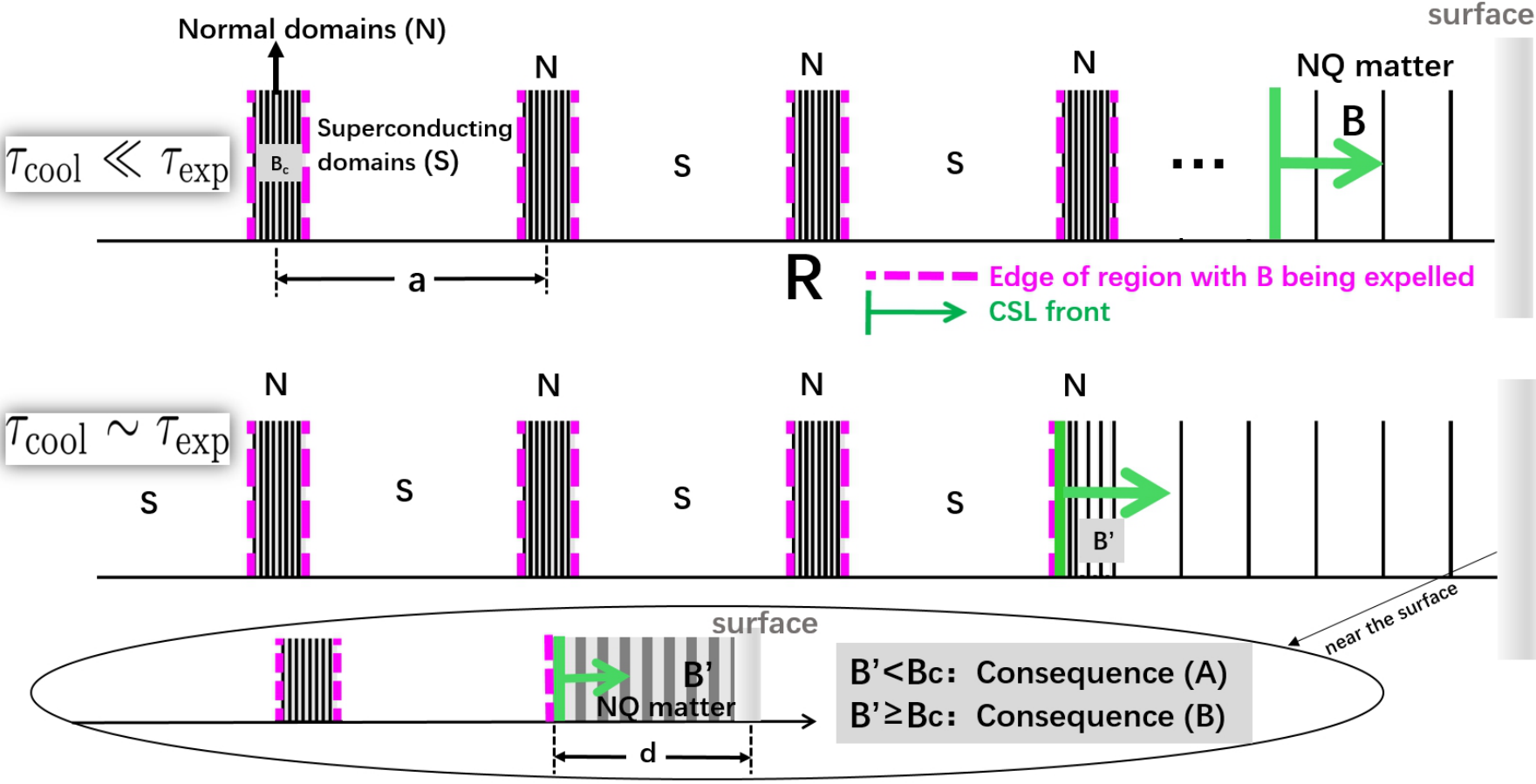}\put(-25,200){(a)}
\\
 \includegraphics[width=.85\textwidth]{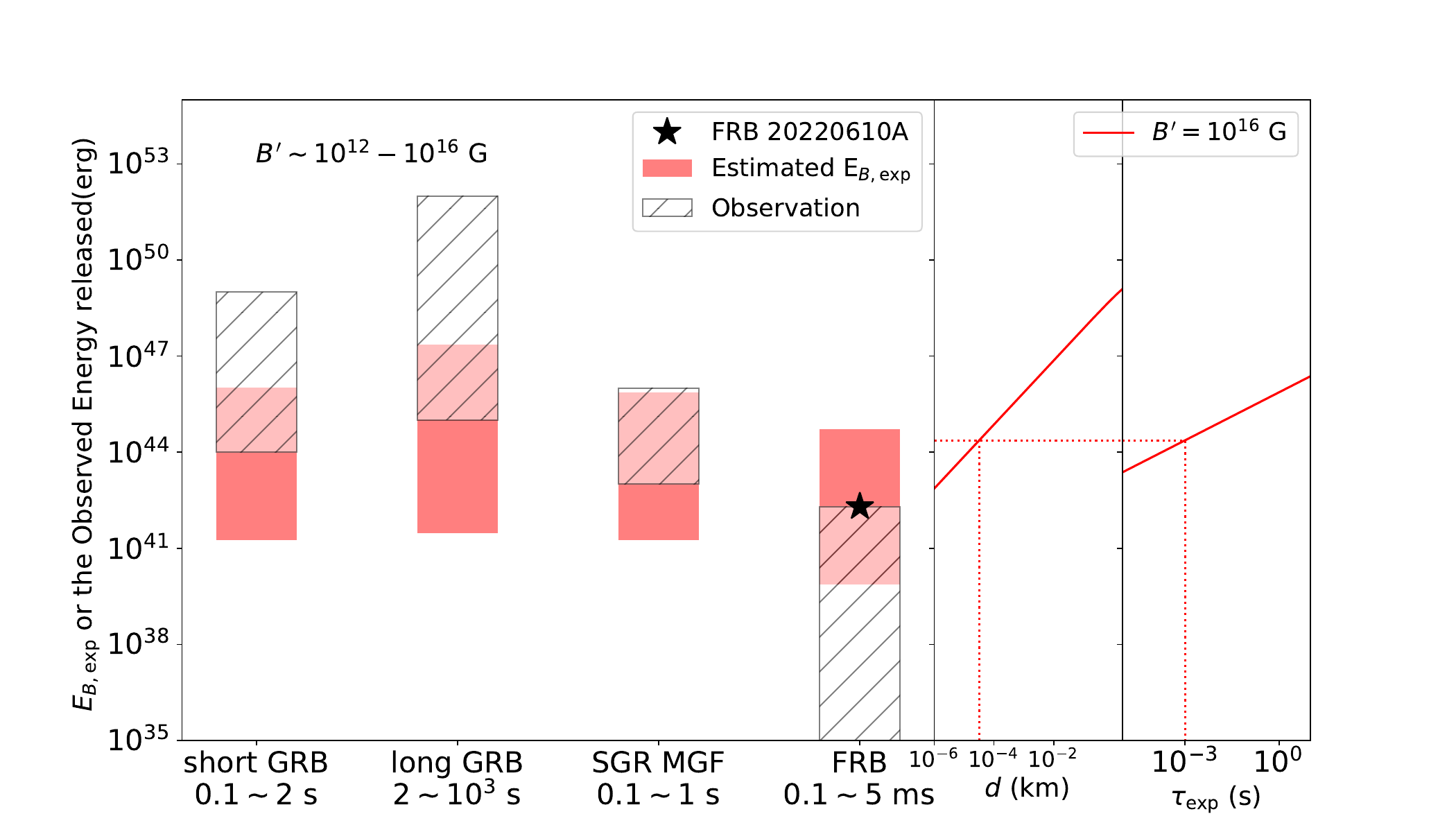}\put(-80,140){(b)}
 \caption{(a) The sketch for the process of the  expulsion of magnetic flux in the cases of $\tau_{\rm cool}\ll\tau_{\rm exp}$ (the first row) and $\tau_{\rm cool}\sim\tau_{\rm exp}$ (the second row with a  elliptical sub-plot). The green thick vertical lines with arrows denote the CSL fronts. The pink dashed lines denote the edges of regions with $B$ being expelled. The regions filled with sparse vertical solid lines denotes the NQ matter with magnetic field $B$. The blank regions labeled with `S' denote the superconducting domains, while those regions labeled with `N' and filled with dense vertical lines between two dashed lines denote normal regions with critical magnetic field $B_{\rm c}$ (densities of these lines do not represent the strength of magnetic field precisely, but for identification purpose only). 
 The detailed description about the process is in the text, and the dimensions in this figure are distorted for clarity. (b) The comparison between the estimated $E_{B, \rm exp}$ (of $M_s=230$ MeV) in the hypothesis of Consequence (A) and observations from different high-energy emissions. The cases for 230 MeV $<M_s\lesssim$ 280 MeV are similar.}\label{fig:compare}
\end{center}
\end{figure*}

Alternating regions of normal and superconducting matters form during this process as sketched in FIG.~\ref{fig:compare} (a). This is the so-called intermediate state of the Type-I superconductor with a configuration consisting of a mixture of normal and superconducting domains ~\citep[][]{1969Natur.224..673B,1974PhR....13..143H,bookCSC}. The expulsion of magnetic flux means that supercurrents form in CSL QM, offset the magnetic field inside to form a superconducting domain and enhance that of outside to keep the total flux a constant. $B^{\prime}$ denotes the magnetic field enhanced in the outside of the superconducting domain. The expulsion around the region containing $B^\prime$ continues until $B^{\prime}= B_{\rm c}$, and a normal domain forms. As the CSL front expands, similar structures form subsequently behind it on a fine scale as shown in the first row in FIG.~\ref{fig:compare} (a). The periodicity length $a$ (distance from center to center between two neighboring normal or superconducting domains) is given by~\cite{1974PhR....13..143H}:
\begin{equation}
    a=\sqrt{\delta \times D/f(\Tilde{B})},
\end{equation}
where $\delta$ denotes the wall-energy parameter, and the wall-energy is associated with the interface separating a normal from a superconducting domain; 
$D$ denotes the thickness of the superconductor placed perpendicularly in a magnetic
field. In the case of $\tau_{\rm cool}\ll\tau_{\rm exp}$, the CSL front quickly reaches the surface, thus $D\sim R$ and has an order of 10 km. $f(\Tilde{h})$ is a numerical function of the reduced field $\Tilde{B} = B/B_{\rm c}$; it reaches the maximum of 0.023 with $\Tilde{B}\simeq0.5$ (see Fig. 2.6 in \citep{bookCSC}).
$\delta$ is given by,
\begin{equation}
    \delta\equiv\xi-\lambda,
\end{equation}
where $\xi$ is the coherence length,  which is approximately equal to the inverse of the gap, i.e. $\xi\sim1/\Delta$; for given $\Delta\sim 1$ MeV in CSL phase, $\xi\simeq 200$ fm; $\lambda$ denotes the penetration depth for the magnetic field in CSL QM and has the order of fm~\citep[][]{2003PhRvL..91x2301S}, thus one has $\lambda\ll\xi$ and $\delta\sim \xi$. With given $B\sim B_{\rm c}/2$  ($B_{\rm c}$ has the order of $10^{16}$ G~\citep{2003PhRvL..91x2301S}), $a$ is estimated to be about $10^{-7}$ km, which is much smaller than the size of QS. The expulsion of a small fraction of $E_{B}$ from the star occurs only if the surface is located in some superconducting domain, and $a$ could be taken as the thickness of the thin layer containing the expelled $E_{\rm B}$. With Equation~(\ref{eq:EB}), $E_{B}$ in this thin layer ($R-a<r<R$) is estimated to be smaller than $10^{42}$ erg for given $B\sim B_{\rm c}/2$, which accounts a small fraction of the total energy inside.

\subsection{ \textbf{The 2nd Case}: $\tau_{\rm cool}\sim\tau_{\rm exp}$} 

  In this case, the edge of the region with $B$ being expelled almost moves together with the CSL front. Note that the region of enhanced $B^\prime$ forms ahead of CSL front in the NQ matter, as shown in the second row in FIG.~\ref{fig:compare} (a).
 Similar alternating structures form once $B^\prime\geq B_{\rm c}$ is valid locally\footnote{Note that $\tau_{\rm cool}\sim\tau_{\rm exp}$ and $B^{\prime}<B_{\rm c}$ being valid during the whole process is an extreme case; $\tau_{\rm exp}$ for all the magnetic field inside the star is about $3\times10^6$ s for given $B=10^{15}$ G and $d=8$ km. The emission of this long duration, even if it exists, is beyond the scope of high-energy EM emissions discussed in this paper.}. When the CSL front (and the edge of the region with $B$ being expelled) gradually approaches the surface with a distance $d$,
 there would be two possible consequences as shown in the elliptical sub-plot in the second row of FIG.~\ref{fig:compare} (a):
 \begin{itemize}
     \item Consequence (A):  $B^{\prime}<B_{\rm c}$ is valid during the following process; the magnetic flux containing in the layer in range of $R-d<r<R$ is expelled from the star (or CSL core); 
     
     \item Consequence (B): $B^{\prime}\geq B_{\rm c}$,
     thus the expulsion stops.
 \end{itemize}
 
 $\tau_{\rm exp}$ of the expulsion of magnetic flux in NQ matter is estimated as~\citep{1969Natur.224..673B,1969Natur.224..674B},
\begin{equation}\label{eq:tau_exp1}
    \tau_{\rm exp}\simeq \frac{4\pi\sigma_{\rm el}}{c^2}\left(\frac{d}{\pi}\right)^2\left(\frac{B^\prime}{H_{\rm c}}\right)^2,
\end{equation}
where $\sigma_{\rm el}$ is the electrical conductivity, $d$ is the scale of variation of the magnetic
field, and $H_{\rm c}\sim10^{19}$ G is the thermodynamical critical field~\citep{2002PhRvD..66a4015I}. The electron conductivity $\sigma_{e, \rm el}$ is given  by the classical expression,
\begin{equation}
    \sigma_{e, \rm el}=\frac{8\pi\alpha_{e} n_el_e}{m_e \overline{v_e}},
\end{equation}
where the $\alpha_e=\frac{1}{137}$ is the QED fine structure constant; $n_e$ is the electron number density; $l_e$ is the electron mean free path, which could be estimated as $n_e^{-1/3}$; $m_e$ is the electron mass and $\overline{v_e}\simeq\sqrt{\frac{T}{m_e}}$ the average thermal velocity of the electrons. 
Near the surface, $\sigma_{\rm el}$ ranges from $3\times10^{22}$ s$^{-1}$ at $M_s=230$ MeV to $7\times10^{22}$ s$^{-1}$ at $M_s=280$ MeV at $T\sim 1$ MeV.
The quark electric conductivity is given by~\citep{1993PhRvD..48.2916H},
\begin{equation}
\sigma_{q, \rm el}= \left(\alpha_{\rm c}\frac{T}{0.08 ~\rm{MeV}}\right)^{-5/3} \left(\frac{\mu}{300 ~\rm{MeV}}\right)^{8/3}\times10^{19}~\rm{s}^{-1}.
\end{equation}
 $\sigma_{q, \rm el}$ has the order of $10^{17}$ s$^{-1}$, which is much smaller than $\sigma_{e, \rm el}$. Therefore, $\sigma_{\rm el}$ is mainly contributed by electrons in NQ matter.

 For Consequence (A), if it is assumed that $E_{B,\rm  exp }$ could convert to energies of other forms which lead to high-energy EM emissions, the relationship between $\tau_{\rm exp} (d)$ and the duration ($\tau_{\rm dur}$) of the corresponding emission is:
 \begin{equation}
     \tau_{\rm dur} =\tau_{\rm exp} (d) + \tau_{\rm rel},
 \end{equation}
 where $\tau_{\rm rel}$ denotes the possible relaxation time in the production of the emission. Thus, in this hypothesis, $\tau_{\rm exp} (d)\lesssim\tau_{\rm dur}$, and a loose upper limit of $E_{B,\rm  exp }$ could be estimated by $\tau_{\rm dur}$ (as shown in the first panel of FIG.~\ref{fig:compare} (b) in solid red bars).

\section{discussion and summary}\label{sec:discussion}

In Section~\ref{sec:QS}, the expulsion of the magnetic field from the surface of quark core is discussed. 

Some high-energy EM emissions can be excluded or allowed in the hypothesis of Consequence (A). For GRBs, the typical energy of the relativistic outflow is estimated to be $f_{\rm b} E_{\rm iso, \gamma}/\eta_\gamma$, where $f_{\rm b}\simeq \theta^2/2$ of the order of $10^{-4}-10^{-3}$ is the beam factor and $\theta$ is the beam opening angle; $E_{\rm iso, \gamma}$
is isotropic-equivalent radiated energy, ranging from $10^{48}-10^{52}$ erg for short GRBs and $10^{49}-10^{55}$ erg for long GRBs; $\eta_\gamma$ is the radiative efficiency~\citep[][]{2018pgrb.book.....Z}. The total energy of the outflow of GRB has an order of $10^{44}-10^{52}$ erg, as shown in hatched bars in FIG.~\ref{fig:compare} (b). Though there exist overlaps between observations and $E_{B,\rm exp}$, considering $\eta_\gamma<1$, the expulsion of magnetic flux because of the CSL condensate forming discussed in Section~\ref{sec:QS} can not provide enough energy for GRBs.

As suggested in \cite{2001PhRvL..87b1101U}, strange star heating events could be taken as a model for GF of SGRs, with the surface of QS heated by impacts of falling massive cometlike objects. Similarly, the surface of QS could be heated in Consequence (A) via magnetic reconnection if the expelled magnetic field lines cross with external ones. A similar mechanism~\citep{2004A&A...420.1025O} has been suggested to generate SGR; the cooled QS has a CFL core acting as a Type-II superconductor in that work and subsequent bursts result from restructuring of the surface magnetic field because of the vortex dynamics. The typical energy of GF is $10^{43}-10^{46}$ erg~\citep[][]{2008ApJ...680..545M}, which seems larger compared with $E_{B,\rm exp}$ as shown in FIG.~\ref{fig:compare} (b). Thus, even if GFs could be induced by the phase transition to CSL condensate, energy provided by the external magnetic field is necessary.

In some theoretical works (e.g., see~\citep[][]{2018ApJ...852..140W, 2020ApJ...902L..32D}), FRBs have been proposed to originate from starquakes. The expulsion of magnetic flux in a short time could lead to an intense evolution of magnetic fields in the magnetosphere, and triggering starquakes by converting $E_{B,\rm exp}$ to kinematic energies of plasma in the magnetosphere and crust~\citep[][]{2020ApJ...902L..32D,2021ApJ...919...89Y}.  In the hypothesis of Consequence (A), the millisecond-duration corresponds to $d\sim10^{-5}$ km and a upper limit of $E_{B, \rm exp}\sim10^{44}$ erg if $B^{\prime}=10^{16}$ G, which means that the expelled magnetic flux is contributed from a thin layer, as shown in the second and third panel in FIG.~\ref{fig:compare} (b). Typical FRBs exhibit isotropic radiative energies of $10^{35}-10^{42}$ erg with the maximum energy reaching $2\times10^{42}$ erg in FRB 20220610A~\cite{2023Sci...382..294R}, which are well below the upper limit of $E_{B, \rm exp}\sim10^{44}$ erg as shown in FIG.~\ref{fig:compare} (b).


In this paper, we investigate the possibility that compact stars act as hideouts for CSL QM in both the MIT bag model and the NJL model. The NJL model does not support the idea of absolutely stable quark matter as well as the existence of conventional pure QS; thus, hybrid configurations with CSL QM in quark cores are constructed; however, it is unstable above $2M_\odot$. The stable configuration of massive QS as the hideout for CSL QM could be reproduced in the MIT bag model. A possible special scenario is proposed in which a small fraction of the magnetic field is expelled from a thin layer near the surface of quark core during the phase transition to the CSL state, while most of $E_{B}$ is frozen inside the star. In conclusion, our analysis suggests that the forming of CSL condensate in quark core could act as an inducement mechanism in some special cases, to power typical FRBs, but as a single source of energy, it is unlikely to generate other EM emissions such as GRBs and GFs. 

Note that the above conclusion is based on the theoretical hypothesis that the conventional QS actually exists. Otherwise, this mechanism can hardly cause these EM emissions of our interest even if the hybrid star could be the hideout for CSL QM. The crusts made of nuclear matter may further affect the final launching of the EM signals. Let us take for example the case of hybrid stars of $P_{\rm NM\rightarrow QM}$= 90 MeV fm$^{-3}$ in the NJL model. The neutron star matter (with a density $\rho$) has an electric conductivity which is estimated to be~\cite[][]{1969Natur.224..673B,1969Natur.224..674B}
\begin{equation}
 \sigma_{\rm NSM}\sim1.5\times10^{23}\left(\frac{8.3~\rm {MeV} }{T}\right)^2 \left(\frac{\rho}{10^{13}~\rm{g}~\rm{cm}^{-3}}\right)^{3/2}\rm{s^{-1}}.
\end{equation}
 $\tau_{\rm exp}$ from the surface of quark core to that of the star could be estimated by integrating $\delta\tau_{\rm exp}$ over the thickness of the crust (about a few km). For given $T\sim1$ MeV and $B=10^{15}$ G, it is found that the time scale is quite enlarged to $10^{10}-10^{11}$ s. Namely, the effect of the expulsion process is enclosed inside the star, so it cannot induce or launch any EM signals of our interest in this paper. 


\acknowledgements
The author thanks the support of the National Natural Science Foundation of China (grant No. 12303052). 
 The author is very grateful for the suggestions and comments from Prof. Shuang-Nan Zhang. The author thanks Prof. Shuang-shi Fang for polishing the language in some paragraphs. I am very grateful for the comments and suggestions from the anonymous referees.
 \appendix
\section{Evaluation of the dispersion relations for spin-0 and CSL condensate}\label{sec:dispersion}

At nonzero temperature, the Matsubara imaginary time formalism is used. The energy $k_0$ is replaced with $-i \omega_n$ where $\omega_n\equiv (2n+1)\pi T$ are the fermionic Matsubara frequencies. The determinant of the inverse quark propagator can be decomposed as follows:
\begin{equation}
   \det \frac{S^{-1}}{T}
=  \prod_{i} 
   \left( \frac{ \omega_n^2 + \epsilon_i (\fettu{k})^2 }{T^2}
   \right)^2 \; .
\label{det-S}
\end{equation}
The Matsubara summation in Equation~(\ref{eq:Omega_s0}) can then be done analytically 
by employing the relation, 
\begin{equation}
\sum_n \ln \left( \frac{ \omega_n^2 + \epsilon_i^2}{T^2} \right) 
= \frac{| \epsilon_i|}{T} 
+ 2 \ln \left( 1 + \e^{- \frac{| \epsilon_i |}{T}} \right) \; .
\end{equation}
Thus, we need to calculate the dispersion relations of $\epsilon_i$ corresponding to zeros of $\det S^{-1}$, i.e. the values
of the energy at which the propagator diverges,
\begin{equation}
\det S^{-1}(\epsilon_i(\fettu{k}), \fettu{k})=0.
\end{equation}

 For the spin-zero case, the off-diagonal components of the propagator (\ref{off-d}) 
are the so-called gap matrices given in terms of three diquark
condensates. The color-flavor structure of these matrices is 
given by
\begin{equation}
\left(\Phi^-\right)^{\alpha\beta}_{ab} 
= -\sum_c \epsilon^{\alpha\beta c}\,\epsilon_{abc}
   \,\Delta_c\,\gamma_5~,
\label{Phim}
\end{equation}
and $\Phi^+ = \gamma^0 (\Phi^-)^\dagger \gamma^0$. Here, as before, 
$a$ and $b$ refer to the color components and $\alpha$ and $\beta$ 
refer to the flavor components. Hence, the gap parameters $\Delta_1$, 
$\Delta_2$, and $\Delta_3$ correspond to the down-strange, the up-strange 
and the up-down diquark condensates, respectively. All three of them 
originate from the color-antitriplet, flavor-antitriplet diquark 
pairing channel. $S^{-1}$ is a $72\times72$ (color: 3, flavor: 3, Dirac: $4\times2$) matrix. With a proper ordering of its rows and columns, 
it decomposes into $6$ diagonal blocks of dimension $4 \times 4$ and one 
diagonal block of dimension $12 \times 12$. The explicit form of 
these blocks
reads
\begin{subequations} 
\begin{equation}
\label{eq:Mp1}
\mathcal{M}_{+}^{(1)}=  
\left( 
\begin{array}{cccc}
 - \mu_d^r + M_d & k & 0 & - \Delta_3 \\
k &  - \mu_d^r - M_d & \Delta_3 & 0 \\
0 & \Delta_3 & \mu_u^g + M_u & k \\
- \Delta_3 & 0 & k & \mu_u^g - M_u
\end{array}
\right) \; ,
\end{equation}
\begin{equation}
\label{eq:Mp2}
\mathcal{M}_{+}^{(2)}=  
\left( 
\begin{array}{cccc}
 \mu_d^r - M_d & k & 0 & - \Delta_3 \\
k &  \mu_d^r + M_d & \Delta_3 & 0 \\
0 & \Delta_3 & -\mu_u^g - M_u & k \\
- \Delta_3 & 0 & k & -\mu_u^g + M_u
\end{array}
\right) \; ,
\end{equation}
\begin{equation}
\label{eq:Mp3}
\mathcal{M}_{+}^{(3)}=  
\left( 
\begin{array}{cccc}
 - \mu_s^r + M_s & k & 0 & - \Delta_2 \\
k &  - \mu_s^r - M_s & \Delta_2 & 0 \\
0 & \Delta_2 & \mu_u^b + M_u & k \\
- \Delta_2 & 0 & k & \mu_u^b - M_u
\end{array}
\right) \; ,
\end{equation}
\begin{equation}
\label{eq:Mp4}
\mathcal{M}_{+}^{(4)}=  
\left( 
\begin{array}{cccc}
 \mu_s^r - M_s & k & 0 & - \Delta_2 \\
k &  \mu_s^r + M_s & \Delta_2 & 0 \\
0 & \Delta_2 & -\mu_u^b - M_u & k \\
- \Delta_2 & 0 & k & -\mu_u^b + M_u
\end{array}
\right) \; ,
\end{equation}
\begin{equation}
\label{eq:Mp5}
\mathcal{M}_{+}^{(5)}=  
\left( 
\begin{array}{cccc}
 - \mu_s^g + M_s & k & 0 & - \Delta_1 \\
k &  - \mu_s^g - M_s & \Delta_1 & 0 \\
0 & \Delta_1 & \mu_d^b + M_d & k \\
- \Delta_1 & 0 & k & \mu_d^b - M_d
\end{array}
\right) \; ,
\end{equation}
\begin{equation}
\label{eq:Mp6}
\mathcal{M}_{+}^{(6)}=  
\left( 
\begin{array}{cccc}
 \mu_s^g - M_s & k & 0 & - \Delta_1 \\
k &  \mu_s^g + M_s & \Delta_1 & 0 \\
0 & \Delta_1 & -\mu_d^b - M_d & k \\
- \Delta_1 & 0 & k & -\mu_d^b + M_d
\end{array}
\right) \; ,
\end{equation}
and
\begin{widetext}
\begin{equation}
\label{eq:Mp7}
\mathcal{M}_{+}^{(7)} =
\left(
\begin{array}{@{\extracolsep{-3.5mm}}cccccccccccc}
-\mu_u^r -M_u &       k      &       0      &      0        &
       0      &       0      &       0      &   -\Delta_3   &
       0      &       0      &       0      &   -\Delta_2   \\
       k      & -\mu_u^r+M_u &       0      &       0       &
       0      &       0      &   \Delta_3   &       0       &
       0      &       0      &   \Delta_2   &       0       \\
       0      &       0      & \mu_u^r -M_u &       k       &
       0      &   \Delta_3   &       0      &       0       &
       0      &   \Delta_2   &       0      &       0       \\
       0      &       0      &       k      & \mu_u^r +M_u  &
   -\Delta_3  &       0      &       0      &       0       &
   -\Delta_2  &       0      &       0      &       0       \\[2mm]
       0      &       0      &       0      &   -\Delta_3   &
\quad 
-\mu_d^g -M_d &       k      &       0      &      0        &
       0      &       0      &       0      &   -\Delta_1   \\
       0      &       0      &   \Delta_3   &       0       &
       k      & -\mu_d^g+M_d &       0      &       0       &
       0      &       0      &   \Delta_1   &       0       \\
       0      &   \Delta_3   &       0      &       0       &
       0      &       0      & \mu_d^g -M_d &       k       &
       0      &   \Delta_1   &       0      &       0       \\
   -\Delta_3  &       0      &       0      &       0       &
       0      &       0      &       k      & \mu_d^g +M_d  &
   -\Delta_1  &       0      &       0      &       0       \\[2mm]
       0      &       0      &       0      &   -\Delta_2   &
       0      &       0      &       0      &   -\Delta_1   &
\quad 
-\mu_s^b -M_s &       k      &       0      &      0        \\
       0      &       0      &   \Delta_2   &       0       &
       0      &       0      &   \Delta_1   &       0       &
       k      & -\mu_s^b+M_s &       0      &       0       \\
       0      &   \Delta_2   &       0      &       0       &
       0      &   \Delta_1   &       0      &       0       &
       0      &       0      & \mu_s^b -M_s &       k       \\
   -\Delta_2  &       0      &       0      &       0       &
   -\Delta_1  &       0      &       0      &       0       &
       0      &       0      &       k      & \mu_s^b +M_s 
\end{array}
\right) .
\end{equation}
\end{widetext}
\end{subequations}
The eigenvalues of $\mathcal{M}_{+}^{(1)}$ and $\mathcal{M}_{+}^{(2)}$ for $gu-rd$ pair are derived as:
\begin{equation}\label{eq:ei1}
\epsilon^{(1)}_{i}=sign_1 \sqrt{\left(\frac{\mu_d^r+\mu_u^g}{2}
sign_2\sqrt{M^2+k^2}\right)^2+\Delta_3^2}
-\frac{\mu_d^r-\mu_u^g}{2}\; ,
\end{equation}
and 
\begin{equation}\label{eq:ei2}
\epsilon^{(2)}_{i}=sign_1 \sqrt{\left(\frac{\mu_d^r+\mu_u^g}{2}
sign_2\sqrt{M^2+k^2} \right)^2+\Delta_3^2}
+\frac{\mu_d^r-\mu_u^g}{2}\; ,
\end{equation}
where two signs, $(sign_1, sign_2)$ of four eigenvalues ($i=1-4$) for each matrix are ($\pm$, $\pm$) and ($\pm$,  $\mp$). Note that we have use the approximation that $\overline{M}=(M_u+M_d)/2$, where the correction on the eigenvalues is 
\begin{equation}
\delta \epsilon=\pm e(\frac{x}{2e^2}-\frac{x^2}{8e^4} +\mathcal{O}(\frac{x^3}{e^6})),
\end{equation}
 $x =M_{u(d)}^2-\overline{M}^2$ and $e=\sqrt{k^2+M^2}$, with  the plus sign for the antiparticle mode and the minus sign for the particle mode. Note the last term in Eq~(\ref{eq:ei1}) or (\ref{eq:ei2}) denotes the half of difference of the quark chemical potential between $u$ and $d$, $\pm(\mu_d-\mu_u)/2$. In the evaluation of $x$, $M_d$ is for the case of plus sign of $\mu_d-\mu_u$ in the last term in the formula of absolute value of the eigenvalue, while $M_u$ is for the case of negative one. The formulae of eigenvalues for $\mathcal{M}_{+}^{(3, 4)}$ for $rs-bu$  and $\mathcal{M}_{+}^{(5, 6)}$ for $gs-bd$  are similar.

For the CFL phase ($\mu>457$ MeV), the approximations of $\Delta_1=\Delta_2$ and $M_u=M_d$ are valid. 12 eigenvalues for $\mathcal{M}_{+}^{(7)}$ are calculated with MATHEMATICA. 4 eigenvalues are for $ru-gd$ pair, which have similar forms to  Eqs~(\ref{eq:ei1}) and (\ref{eq:ei2}), while the rest eight values are much more complex, which are not presented here.  

for the CSL case,  for each flavor, one has
\begin{equation}
\Phi^-
=  \Delta_f (\gamma_3\lambda_2+ \gamma_1\lambda_7+ \gamma_2\lambda_5).
\label{Phim2}
\end{equation} With the similar procedure, dispersion relations for $S_f^{-1}$ can be derived. The flavor structure is trivial for CSL phase, and decoupled from the matrix. For each flavor, $S_f^{-1}$ is a $24\times24$ (color: 3,  Dirac: $4\times2$) matrix. 
The inverse propagator for flavor $f$ is
\begin{equation}
S^{-1}_f(p)=
\left(
\begin{array}{cc}
 \not\!p +\mu_f\gamma^0-M_f&
\hat\Delta_f\\
-{\hat\Delta_f}^\dagger&
\not\!p -\mu_f\gamma^0-M_f
\end{array}~.
\right)
\end{equation}
$\mu_f$ is the chemical potential for the quark of flavor $f$. 
The six dispersions are~\citep{2005PhRvD..72c4008A}:
\begin{eqnarray}
E_{f;1,2}
&=&\sqrt{
\varepsilon_f^2 + \mu_f^2 + |\Delta_f|^2 \mp 2\sqrt{\mu_f^2\varepsilon_f^2
+|\Delta_f|^2{\vec k\,}^{2}}}~,
\nonumber
\\
E_{f;3,5}^2&=& (\varepsilon_{f}-\mu_{f})^2+
c_{f;3,5}^{(1)}\,|\Delta_{f}|^2 + \dots~,
\nonumber
\\
E_{f;4,6}^2&=& (\varepsilon_{f}+\mu_{f})^2+
c_{f;4,6}^{(1)}\,|\Delta_f|^2 + \dots~,
\label{E_1,2} 
\end{eqnarray}
where 
\begin{eqnarray}
 c_{f;3,5}^{(1)}&=&\frac{1}{2}\left[\,5-\frac{\vec k\,^2}{\varepsilon_f\mu_f}
\pm \sqrt{\left(1-\frac{\vec k\,^2}{\varepsilon_f\mu_f}\right)^2
+8\frac{M_f^2}{\varepsilon_f^2}}\,
\right]~, \nonumber \\
c_{f;4,6}^{(1)}&=&\frac{1}{2}\left[\,5+\frac{\vec k\,^2}{\varepsilon_f\mu_f}
\pm \sqrt{\left(1+\frac{\vec k\,^2}{\varepsilon_f\mu_f}\right)^2
+8\frac{M_f^2}{\varepsilon_f^2}}\,
\right]~.  \nonumber \\   
\end{eqnarray}
In the expansions for $E_{f;3,4,5,6}$, higher orders of $\Delta_f$, e.g. $\Delta^4_f$ are neglected in the calculation.

Let us derive the Eq~ (\ref{eq:OmegaCSL_MIT}) in MIT model. 
Since $\Delta_{\rm CSL}$ is very small, around the Fermi surface, $E_{f;1}$ is:
\begin{equation}
   E_{f;1}\simeq \Delta_f\sqrt{1-\frac{\vec k^2}{\mu^2}}=\Delta_f M_f/\mu, 
\end{equation}
Thus the integration for $E_{f;1}$ on the Fermi surface is approximately $\frac{\Delta_f^2 M_f(\mu_f^2-M_f^2)}{2\pi^2\mu_f}.$  $E_{f;3,5}$ around the Fermi surface is 
\begin{equation}
E_{f;3,5}\simeq \Delta_f\sqrt{2\pm\sqrt{2}\frac{M_f}{\mu_f}},
\end{equation}
which is from Eq (31) in \cite{2005PhRvD..72c4008A}; the signs `$+$' and `$-$' correspond to  $E_{f;3}$ and $E_{f;5}$ respectively as derived in \cite{2005PhRvD..72c4008A}. Therefore, the contribution to the thermodynamical potential from CSL condensate is 
\begin{equation}
   \frac{\Delta_f^2(\mu_f^2-M_f^2)}{2\pi^2} \left[\frac{M_f}{\mu_f}+\sum_{sign=+,-}\sqrt{2sign\sqrt{2}\frac{M_f}{\mu_f}}\right] +O(\Delta^4).
\end{equation}
For $u$ and $d$ quarks, their masses could be neglected and contributions are 
\begin{equation}
 \frac{\sqrt{2}}{2\pi^2}\sum_{i=u, d}\Delta_{i}^2\mu_i^2,   
\end{equation}
while for $s$ quark, the contribution is
\begin{equation}
\frac{\Delta_{s}^2(\mu_s^2-M_s^2)}{2\pi^2}\left(\frac{M_s}{\mu_s}+ \sum_{sign=+,-}\sqrt{2 sign\sqrt{2}\frac{M_s}{\mu_s}}\right).
\end{equation}
Then, Eq~(\ref{eq:OmegaCSL_MIT}) is obtained; $\Delta_{\rm CSL}$ in Eq~(\ref{eq:OmegaCSL_MIT}) is redefined as $\Delta_f/\sqrt{2}$, which does not change the order of magnitude of $\Delta_f$.  Note that $E_{f;2,4,6}$ correspond to antiparticles whose contributions could be ignored at very low temperature~\cite{2005PhR...407..205B}. Moreover, in fact, in works~\cite{1984PhRvD..30.2379F,2001PhRvD..64g4017A,2002JHEP...06..031A,2005ApJ...629..969A} and Eq~(\ref{eq:NQ}), antiparticles are not considered either.
\bibliographystyle{apsrev4-1}
\bibliography{hybridCSL}
\end{document}